\documentclass[10pt]{article}
\usepackage{appendix}
\usepackage{setspace}
\usepackage[font=small,  margin=2cm]{caption}
\usepackage[tbtags]{amsmath}
\usepackage{amsthm,amssymb,amsfonts}
\usepackage[numbers]{natbib}
\usepackage{multirow}
\usepackage{lscape}
\usepackage{subfigure}
\usepackage{makecell}
\usepackage{exscale}
\usepackage{booktabs}
\usepackage{array}
\usepackage{fullpage}
\usepackage{url}
\usepackage{algorithm}
\usepackage{algpseudocode}
\usepackage{bm}
\usepackage{smile}
\usepackage{mathtools}
\usepackage{wrapfig}
\usepackage{lipsum}
\usepackage{mathrsfs}
\usepackage{relsize}
\usepackage{dsfont}
\usepackage{multirow}
\usepackage{apalike}
\usepackage{chngcntr}
\usepackage[usenames,dvipsnames,svgnames,table]{xcolor}

\numberwithin{equation}{section}
\numberwithin{theorem}{section}
\numberwithin{corollary}{section}
\counterwithout{asmp}{section}
\numberwithin{definition}{section}

\begin{document}

\title{\LARGE Robust Factor Number Specification  for Large-dimensional  \\ Factor Model}

	\author{Long Yu\thanks{ School of Management, Fudan University, Shanghai, China},~~Yong He\thanks{ School of Statistics, Shandong University of Finance and Economics, Jinan, China; Email:{\tt heyong@sdufe.edu.cn}.},~~ Xinsheng Zhang\thanks{ School of Management, Fudan University, Shanghai, China}}	
	\date{}	
	\maketitle
The accurate specification of the number of factors is critical to the validity of factor models and the topic almost occupies the central position in factor analysis. Plenty of estimators are available under the restrictive condition that the fourth moments of the factors and idiosyncratic errors are bounded. In this paper we  propose efficient and robust estimators for the factor number via considering a more general static Elliptical Factor Model (EFM) framework. We innovatively propose to exploit the multivariate Kendall's tau matrix, which captures the correlation structure of elliptical random vectors. Theoretically we show that the proposed estimators are consistent without exerting any moment condition when both cross-sections ($N$) and time dimensions ($T$) go to infinity. Simulation study shows that the new estimators perform much better in heavy-tailed data setting while performing comparably with the state-of-the-art methods in the light-tailed Gaussian setting. At last, a real macroeconomic data example is given to illustrate its empirical advantages and usefulness.

\vspace{2em}


\section{Introduction}
	Factor models provide a flexible way to extract main features and summarize information from large datasets with relatively smaller number of common factors,  and are wildly applied in research areas such as  finance and biology. A fundamental topic is to consistently determine the number of  latent factors in large-dimensional settings, where cross-sections ($N$) and time dimensions ($T$) go to infinity simultaneously. Plenty of literatures have focused on this topic for  static, dynamic and continuous-time factor models,  including (but not limited to)  \cite{Ahn2013Eigenvalue, 2017ait, Alessi2010Improved, Amengual2007Consistent, Bai2002Determining, Bai2007Determining, 2014caner, Hallin2007Determining, 2017kong, lam2012factor, li2017identifying, Onatski2009Testing, Onatski2010DETERMINING, wu2016robust, Xia2017Transformed, xia2015consistently}.

	Chamberlain and Rothschild  \cite{Chamberlain1983Arbitrage} proposed the static approximate factor models, from which the factor number is often assumed to be known in advance rather than determined by the data until Bai and Ng \cite{Bai2002Determining} first presented consistent estimators for the number of common factors in the large-dimensional setting. The proposed information criteria borrow idea from Akaike information criterion (AIC) while with the penalty term specified as a function of both time dimensions  $T$ and cross-sections $N$.  Alessi {\rm et al.} \cite{Alessi2010Improved} added a tuning parameter on the penalty of the criteria in \cite{Bai2002Determining} and improved stability in the finite samples case as well as the case with large idiosyncratic disturbances. Another line of research on determining factor number mainly relies on the random matrix's eigenvalue theory. Onatski \cite{Onatski2010DETERMINING} provided simple algorithms based on the empirical distribution of the sample covariance matrix's eigenvalues. Lam and Yao \cite{lam2012factor} and Ahn and Horenstein \cite{Ahn2013Eigenvalue} proposed the eigenvalue-based ratio-type estimators separately and independently, which remain reliable even when the idiosyncratic errors are cross-sectionally dependent and serially correlated. Xia et al. \cite{Xia2017Transformed} further improved these estimators by transformation and shrinkage, resulting in better performance in scenarios when  weak, strong or dominated factors exist. The list of literature here is  only illustrative rather than comprehensive.
	
\begin{figure}[hbpt]
	\centering
	\includegraphics[width=8cm,height=6cm]{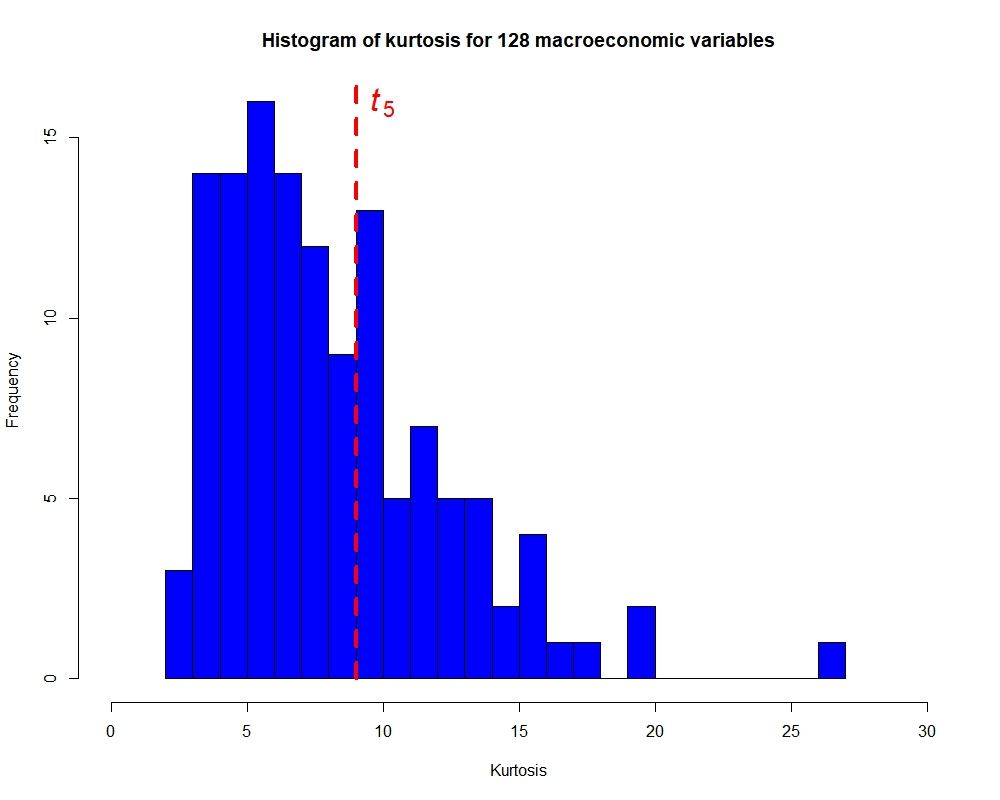}
	\caption{Histogram of the sample kurtosis for 128 macroeconomic variables}
	\label{fig:1}
	\end{figure}

	Both the information-criterion methods and the eigenvalue-based methods  perform well only when some moment constraints are satisfied. The  literatures mentioned above all assume that the fourth moments of common factors and idiosyncratic  errors are bounded.  However, in real data application, it is often the case that we are encountered with  heavy-tailed data and the bounded fourth moment constraints are not satisfied, especially in the areas of finance and economics. Figure \ref{fig:1}  shows the frequency histogram of the sample kurtosis for 128 macroeconomic variables. The data set was originally provided in \cite{McCracken2015FRED},  named as FRED-MD, and is updated to more recent date (from 1959/01 to 2018/02). After removing the time trend, over 1/3 of the 128 variables show larger sample kurtosis than 9, which is the theoretical kurtosis of $t_5$ distribution. Thus it is more reasonable to model  the macroeconomic variables with heavier-tailed distributions such as  $t$ distribution.

\begin{figure}[hbpt]
	\centering
	\includegraphics[scale=0.35]{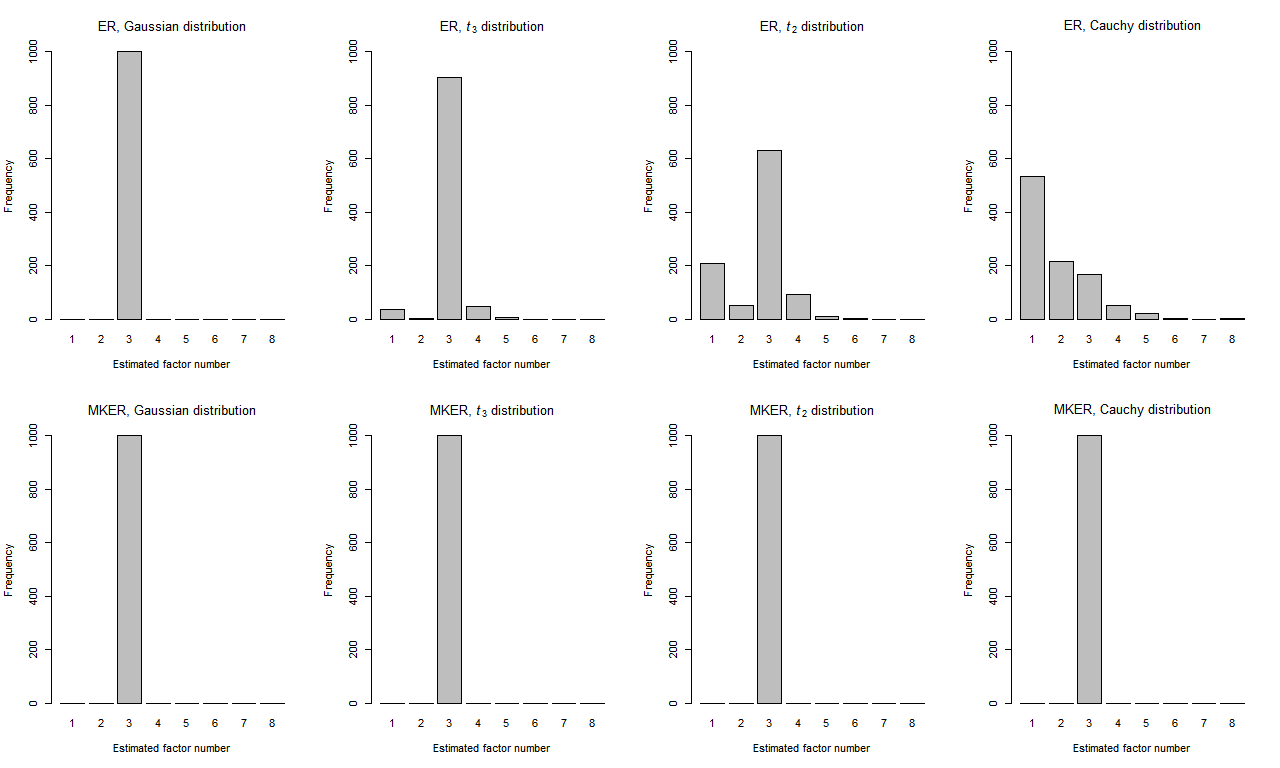}
	\caption{The frequency of the estimated factor number by ``ER" and ``MKER" based on 1000 replications with the true number of factors being 3.}
	\label{fig:2}
	\end{figure}

	Figure \ref{fig:2} further demonstrates the vital importance of taking heavy-tailed feature into consideration when determining the factor number. The empirical performances of two methods are compared, one is the ``ER" method proposed by   \cite{Ahn2013Eigenvalue}, and the other is a modified version named ``MKER" proposed by us. In this example, the true number of factors is 3 and the detailed data-generating  procedure is presented in Section \ref{sec:sim}. Figure \ref{fig:2} shows the barplots for the frequency of the estimated factor number based on 1000 replications. For simulated Gaussian data, ``ER" and ``MKER"  both perform well.  However, for simulated heavy-tailed data from $t$ distribution and Cauchy distribution,  ``MKER" still performs well and shows robustness while ``ER" method gradually loses power as the tail becomes heavier.
	
	Recently, some researchers focus on heavy-tailed factor models \citep[see, for example][]{fan2018,Calzolari2018Estimating,kluppelberg2009copula}.
Fan et al. \cite{fan2018} considered Elliptical Factor Models (EFM) for large-scale covariance estimation. Calzolari and Halbleib  \cite{Calzolari2018Estimating} proposed a factor model structure with $\alpha$-stable distributions, and recommended the indirect inference method for parameter estimation. However, both of the above two papers treated the factor number as given. Kl{\"u}ppelberg and Kuhn \cite{kluppelberg2009copula} proposed a testing procedure to determine the number of common factors under the elliptical copula factor model. They primarily focused on correlation structure with fixed $N$ and cross-sectionally uncorrelated errors.  To the best of our knowledge, our work provides
the first method to specify the factor number for heavy-tailed data with large $N$ and $T$.
	
	In this paper, we propose two consistent estimators for the number of common factors in the EFM framework. The advantages of the proposed methods lie in  the following aspects. Firstly, we don't assume any moment constraints, thus the proposed estimators are consistent even when the observations are from heavy tailed distributions such as $t_2$  or Cauchy. Secondly, the theoretical properties are guaranteed in the  large-dimensional setting where the dimension $N$ can be much larger than sample size $T$. Actually, ${\rm min}\{T,N\}\rightarrow \infty$ is sufficient for guaranteeing the consistency of the proposed estimators. Thirdly, the proposed estimators are eigenvalue-based, thus it's convenient to do similar transformations or shrinkages as in \cite{Ahn2013Eigenvalue} and \cite{Xia2017Transformed} to improve their performances  when  weak, strong or dominated factors exist.
	
 We introduce the notations adopted throughout the paper. For a real number $a$, denote  $[a]$ as the largest integer smaller than or equal to $a$. Let $I(\cdot)$ be the indicator function. Let ${\rm diag}(a_1,\ldots,a_p)$ be a $p\times p$ diagonal matrix, whose diagonal entries are $a_1\ldots,a_p$. It also holds when $\ba_i, i=1,\ldots,p$ are square matrices.  For a matrix $\Ab$, let $\mathrm{A}_{ij}$ (or $\mathrm{A}_{i,j}$) be the $i$-th row, $j$-th column entry of $\Ab$, and let $\Ab^\top$ be the transpose of $\Ab$ and ${\rm Tr}(\Ab)$ be the trace of $\Ab$. Denote $\lambda_j(\Ab)$ as the $j$-th largest eigenvalue of a nonnegative definitive matrix $\Ab$ and let $\|\Ab\|$ be the spectral norm of matrix $\Ab$, $\|\Ab\|_F$ be the Frobenius norm of $\Ab$. For a  nonnegative definite matrix $\Ab$, $\|\Ab\|=\lambda_1(\Ab)$.  For two random variable series $X_n$ and $Y_n$, $X_n\asymp Y_n$ means $X_n=O_p(Y_n)$ and $Y_n=O_p(X_n)$. For two random variables (vectors) $\bX$ and $\bY$, $\bX\stackrel{d}{=}\bY$ means the distributions of $\bX$ and $\bY$ are the same. The constants $c,C_1,C_2$ in different lines can be nonidentical.

	The rest of this paper is organized as follows. In Section \ref{sec:pre}, we introduce the static Elliptical Factor Model (EFM) framework and multivariate Kendall's tau matrix. The construction of the estimators and main theoretical results are shown in Section \ref{sec:mtp}. Section \ref{sec:sim} displays  simulation results and Section \ref{sec:rde} contains a real data example, to empirically illustrate the superiority of the proposed estimators. Conclusions and discussions are provided in Section \ref{sec:con}. Technical proofs and more simulation results are delegated to the Appendix.

	\section{Preliminaries}\label{sec:pre}

	\subsection{Elliptical Distribution}
	The elliptical family contains many frequently-used multivariate distributions such as multivariate Gaussian,  multivariate $t$ distribution. 	We first give the definition of elliptical distribution.
	\begin{definition}[Elliptical Distribution]\label{def2.1}
	We say a random vector $\bX\in \mathbb{R}^d$ has an elliptical distribution, denoted by $\bX\sim EC_d(\bmu,\bSigma,\xi)$, if it has the stochastic representation
	\begin{equation}\label{equ2.1}
	\bX\stackrel{d}{=}\bmu+\xi \Ab \bU,
	\end{equation}
	where $\bmu\in \mathbb{R}^d, \Ab\in\mathbb{R}^{d\times q}, \Ab\Ab^\top=\bSigma\in\mathbb{R}^{d\times d}$, rank$(\bSigma)=q\le d$, $\bU$ is a uniform random vector on the unit sphere in $\mathbb{R}^q$, $\xi$ is a scalar random variable independent of $\bU$ and  $\bSigma=(\Sigma_{ij})$ is  the scatter matrix. The covariance matrix is defined when ${\rm E}(\xi^2)<\infty$, with $\text{Cov}(\bX)={{\rm E}(\xi^2}/{q})\bSigma$.
	\end{definition}
	Definition \ref{def2.1} is unidentifiable from the following two perspectives. First, given a $d\times d$ orthogonal matrix $\bGamma$, define $\tilde \Ab=\Ab\bGamma$, then $\bX\stackrel{d}{=}\bmu+\xi\tilde \Ab\bU$. This is not a vital problem because in most applications  $\bSigma$ is required to be unique rather than the matrix $\Ab$. In this paper, it's sufficient to assume that representation (\ref{equ2.1}) holds for some $\Ab$. Second, if we let $\tilde \xi=c\xi$ and $\tilde\Ab=({1}/{c})\Ab$ for some constant $c$, obviously $\bX\stackrel{d}{=}\bmu+\tilde{\xi}\tilde\Ab\bU$. Fan et al. \cite{fan2018} and Han and Liu \cite{Han2017ECA} assumed that ${\rm E}(\xi^2)=q$ to ensure the identifiability. In this paper, the covariance matrix may be undefined, and we adopt the identifiability condition in \cite{Han2014Scale} that ${\rm max}_i\Sigma_{ii}=1$. The elliptical distribution can also be defined by characteristic function. We only consider continuous elliptical distributions with $\mathbb{P}(\xi=0)=0$.
	
	Some nice properties of Gaussian family still hold for elliptical family. For example, the marginal distributions, conditional distributions and linear combinations of elliptical vectors are all  elliptically distributed. The scatter matrix plays a crucial role in determining the correlation between coordinates of an elliptical vector. In a traditional factor model (light-tailed), the covariance matrix of the large-dimensional vector contains a low-rank common part and the rank is closely related to the number of common factors. Similarly, for elliptical factor model defined later, the scatter matrix contains a low-rank part, which inspires us to focus on the scatter matrix to determine the number of common factors and remove the moment constraints in the conventional methods. The multivariate Kendall's tau matrix is a suitable tool to study the scatter matrices of elliptical vectors, whose definition is given as follows.
	\begin{definition}[Multivariate Kendall's tau Matrix]\label{def2.2}
	For a $d$-dimensional random vector $X$ and its independent copy $\tilde \bX$, the population multivariate Kendall's tau is defined as
	\begin{equation}\label{equ2.2}
	\Kb:={\rm E}\left\{\frac{(\bX-\tilde \bX)(\bX-\tilde \bX)^\top}{\|\bX-\tilde \bX\|^2}\right\}.
	\end{equation}
	Given $n$ independent observations $\bX_1,\ldots,\bX_n$ of $\bX$, the sample multivariate Kendall's tau matrix is
	\begin{displaymath}
	\hat \Kb=\frac{2}{n(n-1)}\sum\limits_{1\le i<j\le n}\frac{(\bX_i-\bX_j)(\bX_i-\bX_j)^\top}{\|\bX_i-\bX_j\|^2}.
	\end{displaymath}
	\end{definition}
	Obviously $\hat\Kb$ is an unbiased estimator of $\Kb$ and $\hat \Kb$ is a matrix-form U-Statistic, with $\|\hat \Kb\|\le 1,\|\Kb\|\le 1$. When $\bX$ is from  elliptical distribution defined by Definition \ref{equ2.1}, it's easy to show $\bX-\tilde \bX\stackrel{d}{=}EC_d(0,\bSigma,\xi_1)\stackrel{d}{=}\xi_1\Ab\bU$, for some random scalar $\xi_1$ determined by $\xi$, where $\bU$ is a uniform random vector on the unit sphere independent with $\xi_1$. By simple calculation, we have
	\begin{displaymath}
	\Kb={\rm E}\left(\frac{\xi_1^2\Ab\bU\bU^\top\Ab^\top}{\|\xi_1\Ab\bU\|^2}\right)=\Ab{\rm E}\left(\frac{\bU\bU^\top}{\|\Ab\bU\|^2}\right)\Ab^\top,
	\end{displaymath}
	which implies that $\xi$ has no effects on $\Kb$. Note that $\xi$ determines whether the moments of elliptical vectors are well defined.  Thus it's possible to relax the moment constraints for estimating the factor number by manipulating with $\Kb$. Han and Liu \cite{Han2017ECA} proved that $\Kb$ shares the same eigenvectors as $\bSigma$, while the relationship between eigenvalues of $\Kb$ and $\bSigma$ are more complicated, see Equation (\ref{equ3.1}) for further details.
	
	{The multivariate Kendall's tau matrix was initially introduced in \cite{choi1998multivariate}, and is also referred as spatial Kendall's tau matrix in the literatures. It has been applied to covariance matrix estimation and principal component analysis in low-dimensional setting (see, for example, \cite{marden1999some,visuri2000sign,croux2002sign}), while Han and Liu \cite{Han2017ECA} and Fan et al. \cite{fan2018} investigated its statistical properties under high-dimensional setting. The multivariate Kendall's tau matrix in  Definition \ref{def2.2} is constructed by ``spatial sign" and should be differentiated from the marginal sign-based Kendall's tau correlation matrix. The definitions of ``spatial sign" and ``marginal sign" can be found in \cite{visuri2000sign}. Recent studies in which marginal Kendall's tau correlation matrix is involved include but are not limited to \cite{mcneil2005quantitative,fang2002meta,lindskog2003kendall,Han2014Scale,He2017High,He2018Variable,He2019Robust}.}
	
	\subsection{Elliptical Factor Model}
	In this paper we focus on the static approximate factor model structure as in \cite{Bai2002Determining}, which has the following expression
	\begin{equation}\label{equ2.3}
	\by_t=\bLambda \bF_t+\bu_t, t=1,\ldots,T,
	\end{equation}
	where $\by_t=(y_{1t},\ldots,y_{Nt})^\top$ are the $N$-dimensional observations, $\bLambda_{N\times r}$ is the unknown factor loading matrix, $\bF_t$ are the $r$-dimensional latent common factors, $\bu_t$ are $N$-dimensional unobservable idiosyncratic random errors. Equation (\ref{equ2.3}) can also be written in matrix form as
	\begin{equation}\label{equ2.4}
	\Yb=\Fb\bLambda^\top+\Ub_{NT},
	\end{equation}
	where $\Yb=(\by_1,\ldots,\by_T)^\top$, $\Fb=(\bF_1,\ldots,\bF_T)^\top$, and $\Ub_{NT}=(\bu_1,\ldots,\bu_T)^\top$.	The research interest is to consistently estimate the latent factor number $r$, which is assumed to be fixed. The specification of $r$  plays an important role in identifying major factors in financial markets. In addition, a suitable choice of $r$ is also fundamental for further factor analysis such as the estimation of common component. It's impossible to identify $\bLambda$ and $\Fb$ from Equation (\ref{equ2.4}) without additional normalization conditions. Either $\bLambda^\top\bLambda/N=\Ib_r$ or $\Fb^\top\Fb/T=\Ib_r$ is frequently used for identification, see, for example,  \cite{Bai2002Determining}. Actually, the identification problem of $\bLambda$ and $\Fb$ has little impact on the specification of $r$. The rank of $\bLambda$ does not change by multiplying an orthogonal matrix or a constant.
	
	Many researchers have focused on this topic since Bai and Ng \cite{Bai2002Determining} gave consistent criteria. But almost all of them exert some moment constraints on $\bF_t$ and $\bu_t$. For example, Bai and Ng \cite{Bai2002Determining} assumed that ${\rm E}\|\bF_t\|^4<\infty$ and ${\rm E}|\bu_t^\top \bu_t/N|\le M$ for all $t=1,\ldots,T$ and some $M>0$, while similar constraints were also found in \cite{Ahn2013Eigenvalue}. In this paper, we aim to extend the results to the more general EFM framework, which relaxes constraints on the moments. We propose to estimate the number of factors based on U-Statistic (the multivariate Kendall's tau matrix) instead of the sample or population moments in existing literature. To this end, additional assumptions are needed and provided as follows. 	
	\begin{asmp}\label{asum:1}
	We assume $(\bF_t^\top,\bu_t^\top)^\top\stackrel{d}{=} EC(\bmu_0,\bSigma_{0},\xi)$, where $\bSigma_0=\left(\begin{matrix}
	\Ib_r&0\\
	0&\bSigma_u
	\end{matrix}\right)$. Further assume $\by_{t}$ are independent observations and $r$ is finite.
	\end{asmp}
	For Assumption \ref{asum:1}, the jointly elliptical distribution of $\bF_t$ and $\bu_t$ entails that $\by_t$ are also elliptically distributed. Similar assumption can be found in \cite{fan2018}. We assume that $\by_t$ are independent, to simplify the technical proof for the convergence of sample multivariate Kendall's tau matrix. Fixed $r$ is a common assumption in related literatures. The scatter matrix corresponding to $\bF_t$ is assumed to be identity matrix, but this can be easily extended to any symmetric positive definite matrix $\bSigma_F$. To this end, take $\bLambda_{\text{new}}=\bLambda\bSigma_F^{1/2}$ and $\bF_{t,\text{new}}=\bSigma_F^{-1/2}\bF_t$.
	\begin{asmp}\label{asum:2}
	We assume that ${C_2}\leq \lambda_m(\bSigma_u)\leq\lambda_1(\bSigma_u)\leq {C_1}$ for positive constants $C_1$ and $C_2$, where  $m={\rm min}\{N,T\}$.
	\end{asmp}	
	Bounded eigenvalues of $\bSigma_u$ in essence make the  idiosyncratic errors negligible compared to the common component. The positive lower bound of $\lambda_m(\bSigma_u)$ can be relaxed similarly to Assumption D in \cite{Ahn2013Eigenvalue} and Assumption C in \cite{Xia2017Transformed}. In other word,  it is sufficient to assume $\lambda_{[d^cm]}(\bSigma_u)\ge c+o_p(1)$ for a positive real number $c$ as well as a real number $d^c\in(0,1]$, which indicates  that asymptotically non-negligible number of eigenvalues from $\bSigma_u$ are lower bounded by a positive real number.
	\begin{asmp}\label{asum:3} There exists a positive definite matrix $\bSigma_{\bLambda}$ with bounded and distinct eigenvalues, i.e.,  $C_2\le\lambda_r(\bSigma_{\bLambda})<\cdots<\lambda_1(\bSigma_{\bLambda})\le C_1$, such that $\|\bLambda^\top \bLambda/N-\bSigma_{\bLambda}\|\rightarrow 0$ as $N\rightarrow\infty$.
	\end{asmp}	
	Assumption \ref{asum:3} assumes $\bLambda^\top \bLambda/N$ converges to a full rank positive definite matrix with bounded maximum and minimum eigenvalues, which is important to the identification of $r$. Otherwise, it's easy to construct new ``mock factors" using linear combinations of the columns of $\Fb$, making $r$ unidentifiable. Besides, by Assumption \ref{asum:3} and the eigenvalue-assumption of $\bSigma_u$ in Assumption \ref{asum:2}, further by Weyl's theorem, the eigenvalues of $\bSigma_{y}$  show the spiked structure in Assumption 2.1 of \cite{fan2018}, where $\bSigma_{y}=\bLambda\bLambda^\top+\bSigma_u$. That is, the spiked eigenvalues $\lambda_1(\bSigma_y),
	\ldots,\lambda_r(\bSigma_y)$ are asymptotically proportional to $N$ while the non-spiked eigenvalues $\lambda_j(\bSigma_y), j>r$ are bounded, i.e., $N(C_2+o(1))\leq \lambda_j(\bSigma_y)\leq N(C_1+o(1))$ for $j\le r$ and $ \lambda_j(\bSigma_y)\leq C_1$ for $j>r$. We assume  $\lambda_j(\bSigma_{\bLambda})$ are distinct, to make corresponding eigenvectors identifiable, which is required in our technical proof.

	\section{Methodology and Theoretical Properties}\label{sec:mtp}
	In this section we present the procedure to estimate the factor number $r$ with the eigenvalues of sample multivariate Kendall's tau matrix defined in Equation (\ref{equ2.2}). The multivariate Kendal's tau shares the same eigenvectors of $\bSigma$, and the eigenvalues show some nonlinear relations with those of $\bSigma$, as stated in \cite{Han2017ECA},
	\begin{equation}\label{equ3.1}
	\lambda_j(\Kb)={\rm E}\left\{\frac{\lambda_j(\bSigma)g_j^2}{\lambda_1(\bSigma)g_1^2+\cdots+\lambda_q(\bSigma)g_q^2}\right\},
	\end{equation}
	where rank$(\bSigma)=q$, $\bg=(g_1,\ldots,g_q)^\top\stackrel{d}{=}{\mathcal{N}(\bm 0,\Ib_q)}$. Equation (\ref{equ3.1}) originates from \cite{marden1999some} and can also be found  in \cite{oja2010multivariate}. As ${\bg}/{\|\bg\|}\stackrel{d}{=}\bU$ with $\bU=(U_1,\ldots,U_q)$ defined in Definition \ref{equ2.1}, $g_j$ in Equation (\ref{equ3.1}) can also be replaced by $U_j$.  Recall that by Assumption \ref{asum:1} and Assumption \ref{asum:2}, the eigenvalues of $\bSigma_y$ show the spiked structure, indicating that we can estimate $r$ with $\lambda_j(\bSigma_y)$,  parallelly similar as the sample covariance matrix eigenvalue-based methods in \cite{Ahn2013Eigenvalue}. Furthermore, by Equation (\ref{equ3.1}), it's possible to estimate $r$ with eigenvalues of $\Kb_y$ or $\hat \Kb_y$ defined as follows:
	\begin{equation}\label{equ3.2}
	\Kb_y={\rm E}\left\{\frac{(\by_1-\by_2)(\by_1-\by_2)^\top}{\|\by_1-\by_2\|^2}\right\},\quad\hat \Kb_y=\frac{2}{T(T-1)}\sum\limits_{1\le s<t\le T}\frac{(\by_s-\by_t)(\by_s-\by_t)^\top}{\|\by_s-\by_t\|^2},
	\end{equation}	
	as long as $\Kb_y$ or $\hat \Kb_y$ also shows a spiked eigenvalue structure. To check this, we need to identify the magnitude of $\lambda_j(\Kb_y),j=1,\ldots,N$. The upper bounds can be easily obtained, i.e.,  $\lambda_j(\Kb_y)\le O(1),j\le r$ and $\lambda_j(\Kb_y)\le o(1), j>r$. To determine $r$, we aim to show that $\lambda_{r+1}(\Kb_y) \ll \lambda_r(\Kb_y) $. The lower bound in \cite{Han2017ECA} is
	\begin{equation}\label{equ3.3}
	\lambda_j(\Kb)\ge\frac{\lambda_j(\bSigma)}{{\rm Tr}(\bSigma)+4\|\bSigma\|_F\sqrt{\ln(N)}+8\|\bSigma\|\ln(N)}\bigg(1-\frac{\sqrt{3}}{N^2}\bigg),
	\end{equation}
which is insufficient to effectively differentiate $\lambda_r(\Kb_y)$ from $\lambda_{r+1}(\Kb_y)$. For illustration, consider a simple case where $\lambda_j(\bSigma_y)=N,j\le r$ and $\lambda_j(\bSigma_y)=c,j>r$ for a positive constant $c$. Then by Inequation (\ref{equ3.3}) we  only get $\lambda_r(\Kb_y)\ge \{c+o(1)\}/{\ln(N)}$, the right hand side of which tends to 0 as $N\rightarrow \infty$. The next lemma shows that with Assumptions \ref{asum:1}-\ref{asum:3}, the asymptotic lower bound shall be a positive constant, i.e., $\lambda_j(\Kb_y)\asymp r^{-1},j\le r$.
	\begin{lemma}\label{lem1}
 Assume Assumptions \ref{asum:1}, \ref{asum:2} and \ref{asum:3} hold, the eigenvalues of the population multivariate Kendall's tau $\Kb_y$ satisfy $\lambda_j(\Kb_y)\asymp r^{-1},j\le r$ {and} $ \lambda_j(\Kb_y)=O(1/m),j>r$, where $m={\rm min}\{N,T\}$.
	\end{lemma}
	By Lemma \ref{lem1}, we are ready for constructing the estimators for factor number $r$. Given observations $\by_1,\ldots,\by_T$,  first get the sample multivariate Kendall's tau matrix by Equation (\ref{equ3.2}), and calculate its eigenvalues $\lambda_j(\hat \Kb_y),j=1,\ldots,N$. We give two estimators motivated separately by the ``ER" method in \cite{Ahn2013Eigenvalue} and the ``TCR" method in \cite{Xia2017Transformed}, which are both eigenvalue-based criteria. Ahn and Horenstein \cite{Ahn2013Eigenvalue} also proposed another estimator called ``GR". Actually ``TCR" is a transformed version of ``GR" with slightly better performance illustrated in \cite{Xia2017Transformed}.

\vspace{1em}

\noindent 	\textbf{Estimator 1: Multivariate {Kendall's tau} Eigenvalue Ratio (MKER)}

\vspace{1em}

Given $\lambda_j(\hat \Kb_y)$ and the possible maximum number of factors $k_{{\rm max}}$, we construct the Multivariate Kendall's tau Eigenvalue Ratio (``MKER") estimator by
	\begin{equation}\label{equ3.4}
	\hat r_{MKER}=\arg{\rm max}_{1\le j\le k_{{\rm max}}}\frac{\lambda_j(\hat \Kb_y)}{\lambda_{j+1}(\hat \Kb_y)}.
	\end{equation}

	To ensure the denominators are not zero, we can add a positive but asymptotically negligible term to each $\lambda_j(\hat \Kb_y)$. Specifically, take $\delta_{NT}=1/\sqrt{m}$, where $m={\rm min}\{N,T\}$, $\hat\lambda_j(\hat \Kb_y)=\lambda_j(\hat \Kb_y)+c\delta_{NT}$ with a small positive constant c and replace $\lambda_j(\hat \Kb_y)$ with $\hat\lambda_j(\hat \Kb_y)$ in Equation (\ref{equ3.4}). The parameter $k_{{\rm max}}$ is a predetermined upper bound of the true factor number $r$. Almost all the existing literatures assume the existence of $k_{{\rm max}}$ to simplify the theoretical proof. Ahn and Horenstein \cite{Ahn2013Eigenvalue} recommended two methods to choose a suitable value for $k_{{\rm max}}$, which are also available for our estimators. In the simulation study, $k_{{\rm max}}$ is set as 8 but this can be replaced with any other reasonable values.

\vspace{1em}

\noindent	\textbf{Estimator 2: Multivariate {Kendall's tau} Transformed Contribution Ratio (MKTCR)}

\vspace{1em}

Let $m={\rm min}\{N,T\}, V_j=\sum_{i=j+1}^{m}\hat\lambda_i(\hat \Kb_y),j=0,\ldots,m-1$, we construct the Multivariate Kendall's tau Transformed Contribution Ratio (``MKTCR") estimator by
	\begin{displaymath}
	\hat r_{MKTCR}=\arg{\rm max}_{1\le j\le k_{{\rm max}}}\frac{\ln\{1+\hat\lambda_j(\hat \Kb_y)/V_{j-1}\}}{\ln\{1+\hat\lambda_{j+1}(\hat \Kb_y)/V_{j}\}}.
	\end{displaymath}	
	
	Similarly, $\hat\lambda_j(\hat\Kb_y)=\lambda_j(\hat\Kb_y)+c\delta_{NT}$ to ensure the denominators are not zero, which is also  important to avoid the case that the ratio goes to infinity for some $j\ne r$.
	
	The reason  why ``ER" and ``MKER" work lie in that the ratio of eigenvalues tends to infinity only when $j=r$. The methods ``TCR" and ``MKTCR" can be regarded as shrinking versions of ``ER" and ``MKER", which eliminate the impact of large or small $\lambda_j(\hat \Kb)$. When {dominant} factors or weak factors exist (corresponding to extremely large or small eigenvalues), ``TCR" and ``MKTCR" may show better finite sample performances by avoiding underestimation of $r$, as stated in \cite{Xia2017Transformed}. When the factors are equally strong or $N,T$ are small, ``ER" and ``MKER" shall be more {accurate and} reliable.
	When $N,T$ go to infinity simultaneously, {all} of them converge to the true number of factors $r$ in probability under some assumptions. The following theorem is the main theoretical result of this paper.
	\begin{theorem}\label{the1}
	Assume Assumptions \ref{asum:1}-\ref{asum:3} hold and $r\ge 1$, then we have
	\begin{displaymath}
	\lim\limits_{m\rightarrow\infty}\Pr(\hat r_{MKER}=r)=1,\quad	\lim\limits_{m\rightarrow\infty}\Pr(\hat r_{MKTCR}=r)=1,\quad \text{for}\quad k_{{\rm max}}\in [r,m-1].
	\end{displaymath}
	\end{theorem}	
	 For the case $r=0$, we can slightly modify the proposed estimators by defining a mock eigenvalue $\hat\lambda_0(\hat \Kb_y)=-1/\ln(\delta_{NT})$ such that $\hat\lambda_0(\hat \Kb_y)\rightarrow0$ and $\hat\lambda_0(\hat \Kb_y)/{\delta_{NT}}\rightarrow\infty$. In detail, define two new estimators in advantage of the mock eigenvalue, as follows

\[
\tilde r_{MKER}=\arg{\rm max}_{0\le j\le k_{{\rm max}}}\frac{\hat\lambda_j(\hat \Kb_y)}{\hat\lambda_{j+1}(\hat \Kb_y)} \quad \text{and} \quad\tilde r_{MKTCR}=\arg{\rm max}_{0\le j\le k_{{\rm max}}}\frac{\ln\{1+\hat\lambda_j(\hat \Kb_y)/V_{j-1}\}}{\ln\{1+\hat\lambda_{j+1}(\hat \Kb_y)/V_{j}\}}.
\]
We have the following corollary which guarantees the consistency of $\tilde r_{MKER}$ and $\tilde r_{MKTCR}$.
	\begin{corollary}\label{cor1}
	With Assumptions \ref{asum:1}-\ref{asum:3}, $r\ge 0$ and $\hat\lambda_0(\hat\Kb_y)$, we have
	\begin{displaymath}
	\lim\limits_{m\rightarrow\infty}\Pr(\tilde r_{MKER}=r)=1,\quad	\lim\limits_{m\rightarrow\infty}\Pr(\tilde r_{MKTCR}=r)=1,\quad \text{for}\quad k_{{\rm max}}\in [r,m-1].
	\end{displaymath}
	\end{corollary}
	\begin{remark}
			The proposed approaches are appealing because we remove the moment constraints required in previous literatures, consequently providing reliable estimators for $r$ even for extremely heavy-tailed data. As one reviewer pointed out, the elliptical distribution assumption is another strong shape constraint, even though elliptical family contains many widely-used distributions. Without this shape  assumption, bounded fourth moment constraint seems to be unavoidable. This is a trade-off  between distribution assumption and moment assumption and it arises for many statistical problems.
		\end{remark}

	\section{Simulation Study}\label{sec:sim}
	In this section, we divide the  simulation study into three {parts} to thoroughly compare the proposed estimators with other competitors. Main competitors we consider are the ``ER" and ``GR" in \cite{Ahn2013Eigenvalue} and ``TCR" in \cite{Xia2017Transformed}. We exclude criteria in \citep{Alessi2010Improved,Bai2002Determining,Onatski2010DETERMINING} because Xia et al. \cite{Xia2017Transformed} concluded that these methods {always} perform no better than ``TCR".
	We use similar data-generating models as in \cite{Ahn2013Eigenvalue} and \cite{Xia2017Transformed}. That is,
	\begin{align}
	&y_{it}=\sum\limits_{j=1}^{r}\lambda_{ij}F_{jt}+\sqrt{\theta}u_{it},\quad u_{it}=\sqrt{\frac{1-\rho^2}{1+2J\beta^2}}e_{it}, \nonumber \\
&e_{it}=\rho e_{i,t-1}+(1-\beta)v_{it}+\sum_{l={\rm max}\{i-J,1\}}^{{\rm min}\{i+J,N\}}\beta v_{lt}, t=1,\ldots,T,i=1,\ldots,N,\nonumber
	\end{align}
	where  $F_{jt}$ and $v_{it}$ are generated from  heavy-tailed distributions in the simulation study. In most of our settings, we set $r=3$ and let $\lambda_{ij}$ be independently drawn from standard normal distribution $\mathcal{N}(0,1)$. The parameter $\theta$ controls the SNR (signal to noise ratio), $\rho$ controls the serial correlations of idiosyncratic errors, while $\beta$ and $J$ control the cross-sectional correlations. We point out that although we  assume $\by_t$ to be independent in Assumption \ref{asum:1}, we consider the serially correlated structure of $\bu_t$ to compare these methods more comprehensively. In the simulation study,  $y_{it}$ are doubly demeaned according to \cite{Ahn2013Eigenvalue}. That is, we apply all the methods to the demeaned data $\tilde y_{it}$ where
\[
\tilde y_{it}=y_{it}-\frac{1}{T}\sum_{q=1}^Ty_{iq}-\frac{1}{N}\sum_{p=1}^Ny_{pt}+\frac{1}{NT}\sum_{p,q}y_{pq}.
\]
	\subsection{Simulation Part I}
	In this part, we use the following data-generating procedure to compare these estimators when data are from diversified population distributions.
	\begin{description}
	\item[Scenario A] Set $r=3,k_{{\rm max}}=8,\theta=1,\rho=\beta=J=0$, $N=T=25,\ldots,200$, $(\bF_t^\top,\bv_t^\top)$ are \emph{i.i.d.} jointly elliptical random vectors.
	\end{description}
	We consider multivariate Gaussian $\mathcal{N}(\bf{0},\bI_{N+r})$ and multivariate centralized $t$ distribution $t_{\nu}(\bf{0},\bI_{N+r})$ with $\nu=3,2,1$. The \emph{p.d.f.} of a $d$-dimensional multivariate $t$ distribution $t_{\nu}(\bmu,\bSigma_{d\times d})$ is
	\begin{displaymath}
	\frac{{\Gamma\big((\nu+d)/2\big)}}{\Gamma(\nu/2)\nu^{d/2}\pi^{d/2}|\bSigma|^{1/2}}\bigg\{1+\frac{1}{\nu}(\bx-\bmu)^\top\bSigma^{-1}(\bx-\bmu)\bigg\}^{-(\nu+d)/2},
	\end{displaymath}
	 {where $\Gamma(\cdot)$ is the gamma function}. Note that when $\nu=1$, it's the multivariate Cauchy distribution. To check how the proposed methods perform when Assumption \ref{asum:1} doesn't hold, we also consider another two cases. For the first case, $(\bF_t^\top,\bv_t^\top)$ are generated from multivariate skew-$t_3$ distributions, consequently the symmetric assumption in Assumption \ref{asum:1} is no longer satisfied. For the second case,  $\bF_t$ and $\bv_t$ are generated independently from multivariate $t_3$ distribution, thus $\by_t$ may not be elliptically distributed. We use the \texttt{rmvt} function in the \textsf{R} package \texttt{mvtnorm} to generate the multivariate $t$ data. Multivariate skew-$t_3$ are generated from $\mathcal{ST}_{N+r}(\bxi={\bf 0},\bOmega=\Ib,\balpha={\bf 20},\nu=3)$ by function \texttt{rmvst} in \textsf{R} package \texttt{fMultivar}. The results are reported  in the form $x(y|z)$ and shown in Table \ref{tab:1}, from small $N,T=25$ to large $N,T=200$, in which  $x$ is the sample mean of the estimated factor number based on 1000 replications, $y$ and $z$ are the numbers of underestimation and overestimation. Figure \ref{fig:2} in the introduction section also illustrate a part of the results with $N=T=125$.

	\begin{table*}[hbpt]
	  	\begin{center}
	  		\addtolength{\tabcolsep}{1pt}
	  		\small
	  		\caption{Empirical results for  Scenario $\Ab$: $r=3,k_{{\rm max}}=8,\theta=1,\rho=\beta=J=0$, $(\bF_t^\top,\bv_t^\top)$ from multivariate elliptical family.``Gaussian" for multivariate Gaussian distribution. ``$t_3$", ``$t_2$" and ``Cauchy"  for multivariate $t$ with degree of freedom 3,2,1. ``Skew $t_3$ for multivariate skew $t$ with degree of 3.  ``Ind $t_3$" for the case where  $\bF_t$ and $\bv_t$ are generated independently from multivariate $t_3$ distribution.}\label{tab:1}
\renewcommand{\arraystretch}{1.20}
	   \scalebox{0.9}{ 		\begin{tabular*}{16.5cm}{lllllllll}
	  			\toprule[1.2pt]
       		Family&$N$&$T$&$r$&$\hat r_{GR}$&$\hat r_{ER}$&$\hat r_{MKER}$&$\hat r_{TCR}$&$\hat r_{MKTCR}$\\\hline
       	    \multirow{8}*{Gaussian}&25&25&3&2.936(60$|$1)&2.822(134$|$0)&2.895(91$|$0)&2.953(45$|$1)&2.953(46$|$0)\\
      	    &50&50&3&3.000(0$|$0)&3.000(0$|$0)&3.000(0$|$0)&3.000(0$|$0)&3.000(0$|$0)\\
       	    &75&75&3&3.000(0$|$0)&3.000(0$|$0)&3.000(0$|$0)&3.000(0$|$0)&3.000(0$|$0)\\
       	    &100&100&3&3.000(0$|$0)&3.000(0$|$0)&3.000(0$|$0)&3.000(0$|$0)&3.000(0$|$0)\\
       	    &125&125&3&3.000(0$|$0)&3.000(0$|$0)&3.000(0$|$0)&3.000(0$|$0)&3.000(0$|$0)\\
       	    &150&150&3&3.000(0$|$0)&3.000(0$|$0)&3.000(0$|$0)&3.000(0$|$0)&3.000(0$|$0)\\
       	    &175&175&3&3.000(0$|$0)&3.000(0$|$0)&3.000(0$|$0)&3.000(0$|$0)&3.000(0$|$0)\\
       	    &200&200&3&3.000(0$|$0)&3.000(0$|$0)&3.000(0$|$0)&3.000(0$|$0)&3.000(0$|$0)\\
       		\hline
       	    \multirow{8}*{$t_3$}&25&25&3&2.686(324$|$117)&2.276(497$|$42)&2.799(162$|$1)&2.844(257$|$151)&2.927(81$|$13)\\
      	    &50&50&3&3.012(86$|$125)&2.726(203$|$56)&3.000(0$|$0)&3.151(40$|$165)&3.000(0$|$0)\\
       	    &75&75&3&3.077(34$|$120)&2.919(91$|$71)&3.000(0$|$0)&3.164(14$|$154)&3.000(0$|$0)\\
       	    &100&100&3&3.092(17$|$106)&2.965(55$|$61)&3.000(0$|$0)&3.162(2$|$142)&3.000(0$|$0)\\
       	    &125&125&3&3.068(19$|$88)&2.968(45$|$52)&3.000(0$|$0)&3.139(5$|$118)&3.000(0$|$0)\\
       	    &150&150&3&3.082(9$|$87)&3.007(26$|$54)&3.000(0$|$0)&3.124(4$|$113)&3.000(0$|$0)\\
       	    &175&175&3&3.102(8$|$98)&3.007(32$|$60)&3.000(0$|$0)&3.144(2$|$122)&3.000(0$|$0)\\
       	    &200&200&3&3.075(4$|$78)&2.994(28$|$47)&3.000(0$|$0)&3.109(0$|$100)&3.000(0$|$0)\\
       		\hline
       	    \multirow{8}*{$t_2$}&25&25&3&2.560(441$|$179)&2.075(619$|$81)&2.778(181$|$11)&2.938(330$|$268)&2.938(86$|$33)\\
      	    &50&50&3&2.859(235$|$181)&2.384(418$|$70)&3.000(0$|$0)&3.156(143$|$256)&3.000(0$|$0)\\
       	    &75&75&3&2.933(193$|$186)&2.523(342$|$86)&3.000(0$|$0)&3.258(96$|$269)&3.000(0$|$0)\\
       	    &100&100&3&3.070(133$|$223)&2.614(285$|$113)&3.000(0$|$0)&3.337(57$|$295)&3.000(0$|$0)\\
       	    &125&125&3&3.082(118$|$228)&2.667(256$|$111)&3.000(0$|$0)&3.345(49$|$306)&3.000(0$|$0)\\
       	    &150&150&3&3.087(113$|$210)&2.710(237$|$115)&3.000(0$|$0)&3.363(36$|$285)&3.000(0$|$0)\\
       	    &175&175&3&3.051(98$|$180)&2.735(209$|$99)&3.000(0$|$0)&3.300(30$|$251)&3.000(0$|$0)\\
       	    &200&200&3&3.164(81$|$221)&2.764(205$|$113)&3.000(0$|$0)&3.393(25$|$298)&3.000(0$|$0)\\
       		\hline
       	    \multirow{8}*{Cauchy}&25&25&3&2.464(572$|$207)&1.869(756$|$91)&2.756(195$|$9)&3.161(387$|$357)&2.914(115$|$42)\\
      	    &50&50&3&2.640(529$|$248)&1.846(748$|$78)&2.994(5$|$0)&3.349(338$|$413)&2.999(1$|$0)\\
       	    &75&75&3&2.647(510$|$250)&1.936(719$|$96)&3.000(0$|$0)&3.528(288$|$458)&3.000(0$|$0)\\
       	    &100&100&3&2.740(495$|$277)&1.889(732$|$91)&3.000(0$|$0)&3.659(268$|$490)&3.000(0$|$0)\\
       	    &125&125&3&2.578(535$|$251)&1.887(735$|$93)&3.000(0$|$0)&3.614(285$|$484)&3.000(0$|$0)\\
       	    &150&150&3&2.679(515$|$257)&1.864(747$|$86)&3.000(0$|$0)&3.672(264$|$489)&3.000(0$|$0)\\
       	    &175&175&3&2.681(525$|$267)&1.835(749$|$82)&3.000(0$|$0)&3.707(269$|$498)&3.000(0$|$0)\\
       	    &200&200&3&2.726(493$|$291)&1.930(726$|$102)&3.000(0$|$0)&3.694(256$|$505)&3.000(0$|$0)\\\hline

  	\multirow{8}*{Skew $t_3$}&25&25&3&2.738(302$|$127)&2.317(470$|$49)&2.833(143$|$4)&2.959(217$|$181)&2.919(84$|$10)
\\
	&50&50&3&2.995(75$|$105)&2.76(175$|$53)&3.000(0$|$0)&3.114(36$|$143)&3.000(0$|$0)
\\
	&75&75&3&3.101(22$|$119)&2.934(80$|$65)&3.000(0$|$0)&3.156(12$|$148)&3.000(0$|$0)\\
	&100&100&3&3.064(19$|$81)&2.979(42$|$47)&3.000(0$|$0)&3.137(7$|$117)&3.000(0$|$0)
\\
	&125&125&3&3.083(11$|$92)&2.999(36$|$62)&3.000(0$|$0)&3.134(4$|$126)&3.000(0$|$0)
\\
	&150&150&3&3.104(10$|$103)&2.990(42$|$64)&3.000(0$|$0)&3.162(1$|$133)&3.000(0$|$0)
\\
	&175&175&3&3.083(10$|$80)&3.009(25$|$49)&3.000(0$|$0)&3.137(3$|$109)&3.000(0$|$0)
\\
	&200&200&3&3.074(13$|$88)&2.996(30$|$51)&3.000(0$|$0)&3.135(1$|$116)&3.000(0$|$0)
\\\hline
	\multirow{8}*{Ind $t_3$}&25&25&3&2.906(315$|$260)&2.471(470$|$164)&2.651(292$|$26)&3.124(246$|$314)&2.94(174$|$96)
\\
	&50&50&3&3.291(76$|$298)&2.985(174$|$218)&2.994(6$|$0)&3.408(52$|$336)&2.995(5$|$0)\\
	&75&75&3&3.358(34$|$296)&3.153(86$|$227)&3.000(0$|$0)&3.452(16$|$327)&3.000(0$|$0)
\\
	&100&100&3&3.311(16$|$254)&3.182(46$|$209)&3.000(0$|$0)&3.394(5$|$287)&3.000(0$|$0)
\\
	&125&125&3&3.288(9$|$223)&3.165(32$|$179)&3.000(0$|$0)&3.354(2$|$249)&3.000(0$|$0)
\\
	&150&150&3&3.330(4$|$246)&3.207(22$|$195)&3.000(0$|$0)&3.373(3$|$266)&3.000(0$|$0)
\\
	&175&175&3&3.246(7$|$207)&3.165(18$|$168)&3.000(0$|$0)&3.315(3$|$241)&3.000(0$|$0)
\\
	&200&200&3&3.304(3$|$224)&3.218(13$|$189)&3.000(0$|$0)&3.350(0$|$249)&3.000(0$|$0)\\
	  	\bottomrule[1.2pt]		
	  	\end{tabular*}}\\
	  \end{center}
	\end{table*}
	
    From Table \ref{tab:1}, we can see that all the five estimators perform quite well under multivariate Gaussian distribution, even with relatively small $N,T=50$. For the heavy-tailed cases, ``MKER" and ``MKTCR" still work well  while ``ER" tends to underestimate and ``TCR" tends to overestimate. The estimated factor number from ``GR" tends to be larger than ``ER" and smaller than ``TCR". Besides, ``GR" is also not effective in heavy-tailed settings. For Cauchy distribution setting, the conventional estimators perform even worse as what's expected.

   An interesting finding is that the proposed estimators still outperform competitors for the two cases where Assumption \ref{asum:1} is no longer satisfied. Even though we can't relax Assumption \ref{asum:1} for theoretical analysis by now, we give an intuitive explanation on the above finding here. By the construction of multivariate Kendall's tau, $\Kb_y$ depends on the distribution of $\by_t-\tilde\by_t$, where $\tilde \by_t$ is an independent copy. Hence, $\by_t-\tilde\by_t$ remains symmetric no matter how $\by_t$ is generated, which may be the reason why our methods still work for asymmetric case. On the other hand, when $\bF_t$ and $\bu_t$ are independently generated from elliptical families, for example,
    \[
    \bF_t\stackrel{d}{=}\xi_1\Ub_1\quad\text{and}\quad \bu_t\stackrel{d}{=}\xi_2\Ub_2,
    \]
    we  have $\by_t\stackrel{d}{=}\xi_1(\Lb\Ub_1+\xi_2\xi_1^{-1}\Ub_2)$. Therefore, even $\by_t$ is not elliptically distributed, the distortion from idiosyncratic errors may still be controlled when $\xi_2/\xi_1$ is not too large.

	\subsection{Simulation Part II}
	In this part, we consider most of the simulation settings in \cite{Xia2017Transformed} to further compare these estimators for Gaussian data. In specific, the following scenarios are considered:
	\begin{description}
	\item[Scenario B1] Serially and cross-sectionally correlated errors: $r=3,\theta=1,\rho=0.5,\beta=0.2,k_{{\rm max}}=8,J={\rm max}\{10, N/20\}$, $(\bF_t^\top,\bv_t^\top)\sim \mathcal{N}(\bf{0},\Ib_{N+r})$, $N=T=25,\ldots,200$.	
	\item[Scenario B2] Weak factors: $r=3,\theta=6,\rho=0.5,\beta=0.2,k_{{\rm max}}=8,J={\rm max}\{10, N/20\}$, $(\bF_t^\top,\bv_t^\top)\sim \mathcal{N}(\bf{0},\Ib_{N+r})$, $N=T=25,\ldots,200$.
	\item[Scenario B3] Strong and weak factors: $r=3,\theta=1,\rho=0.5,\beta=0.2,k_{{\rm max}}=8,J={\rm max}\{10, N/20\},N=T=100$, $(\bF_t^\top,\bv_t^\top)\sim \mathcal{N}(\bf{0},\bD)$, {$\Db$} is $(N+r)\times(N+r)$ diagonal with $\mathrm{D}_{ii}=1,i\ne 3$ and $\mathrm{D}_{33}=SNR$ with $SNR$ from 0.7 to 0.4.
	\item[Scenario B4] Choice of $k_{{\rm max}}$: $r=3,\theta=1,\rho=0.5,\beta=0.2,J={\rm max}\{10, N/20\},N=T=100$, $(\bF_t^\top,\bv_t^\top)\sim \mathcal{N}(\bf{0},\bI_{N+r})$, $k_{{\rm max}}=8,12,16,20,25,30$.
	\item[Scenario B5] Dominant factor:  $r=2,\theta=1,\rho=0.5,\beta=0.2,k_{{\rm max}}=8,J={\rm max}\{10, N/20\},N=T=100$, $(\bF_t^\top,\bv_t^\top)\sim\mathcal{N}(\bf{0},\Db)$, $\Db$ is $(N+r)\times(N+r)$ diagonal with $\mathrm{D}_{ii}=1,i\ne 1;\mathrm{D}_{11}=SNR$ with $SNR$ from 1 to 20.
	\end{description}
	Scenario $\Bb\mathbf{1}$ is a simple case containing serially and cross-sectionally correlated errors with Gaussian distributions. Scenario $\Bb\mathbf{2}$, Scenario $\Bb\mathbf{3}$ and Scenario $\Bb\mathbf{5}$ corresponds to existence of weak factors, strong factors, and dominant factors respectively, which are in favor of the  shrinking estimators ``TCR" and ``MKTCR". In Scenario $\Bb\mathbf{4}$, we consider the impact of the choice of $k_{{\rm max}}$ for different methods. The simulation results totally meet our expectations that ``MKER" and ``MKTCR" perform comparably to ``ER" and ``TCR" in Gaussian cases, though with slightly lower convergence rates. We  show the simulation results of Scenario $\Bb\mathbf{1}$ in Table \ref{tab:2}, and designate the remaining to the Appendix.
\begin{table*}[hbpt]
  	\begin{center}
  		\addtolength{\tabcolsep}{4pt}
  		\small \caption{Simulation results for Scenario $\Bb\mathbf{1}$: $r=3,\theta=1,\rho=0.5,\beta=0.2,k_{{\rm max}}=8,J={\rm max}\{10, N/20\}$, $(\bF_t^\top,\bv_t^\top)\sim \mathcal{N}(\bf{0},\bI_{N+r})$. Effects of serial and cross-sectional correlations.}
  		\vspace{0.2cm}\label{tab:2}
  \renewcommand{\arraystretch}{1.55}
  		\begin{tabular*}{16cm}{llllllll}
  			\toprule[1.2pt]
       		$N$&$T$&$r$&$\hat r_{GR}$&$\hat r_{ER}$&$\hat r_{MKER}$&$\hat r_{TCR}$&$\hat r_{MKTCR}$\\\hline
       	    25&25&3&3.288(42$|$311)&3.013(119$|$164)&3.170(67$|$232)&3.400(26$|$380)&3.457(24$|$399)\\
      	    50&50&3&3.626(1$|$345)&3.199(27$|$146)&3.355(11$|$225)&3.945(0$|$491)&4.103(0$|$561)\\
       	    75&75&3&3.099(1$|$33)&3.006(7$|$4)&3.036(4$|$16)&3.288(0$|$91)&3.465(0$|$139)\\
       	    100&100&3&2.999(1$|$0)&2.999(1$|$0)&2.999(1$|$0)&3.000(0$|$0)&3.014(0$|$3)\\
       	    125&125&3&3.000(0$|$0)&3.000(0$|$0)&3.000(0$|$0)&3.000(0$|$0)&3.000(0$|$0)\\
       	    150&150&3&3.000(0$|$0)&3.000(0$|$0)&3.000(0$|$0)&3.000(0$|$0)&3.000(0$|$0)\\
       	    175&175&3&3.000(0$|$0)&3.000(0$|$0)&3.000(0$|$0)&3.000(0$|$0)&3.000(0$|$0)\\
       	    200&200&3&3.000(0$|$0)&3.000(0$|$0)&3.000(0$|$0)&3.000(0$|$0)&3.000(0$|$0)\\
  	\bottomrule[1.2pt]		
  	\end{tabular*}\\
  \end{center}
\end{table*}

	From Table \ref{tab:2}, we can see that  when $N,T\ge 100$, the proposed methods $\hat r_{MKER}$ and $\hat r_{MKTCR}$ perform as well as $\hat r_{ER}$ and $\hat r_{TCR}$ even with serially correlated errors, which implies a quite fast convergence rate. Together with the simulation results shown in the Appendix, we claim that ``MKER" and ``MKTCR" are effective and show comparable performances with ``ER" and ``TCR" in Gaussian cases, even with serially and cross-sectionally correlated errors. When strong factors, weak factors, or dominant factors exists, ``MKTCR" tends to perform better than ``MKER".  The choice of $k_{{\rm max}}$ shows limited effects on ``MKER" and ``MKTCR" as well as on ``ER" and ``TCR".
	
	\subsection{Simulation Part III}
	In this section we extend the scenarios in Simulation Part II to the heavy-tailed cases. We only  replace $\bF_t$ and $\bv_t$ with some jointly elliptical random vectors. Multivariate $t_3$ distributions are used in the following simulations, and similar results are obtained for  $t_2$ and Cauchy distributions. We denote the corresponding scenarios as Scenarios $\Cb\mathbf{1}$-$\Cb\mathbf{5}$. For each Scenario $\Cb\mathbf{i}$, the parameter settings are set the same as Scenario  $\Bb\mathbf{i}$,  except that $(\bF_t^\top,\bv_t^\top)\sim t_3(\bf{0},\bSigma)$ with the scatter matrices $\bSigma$ equal to the covariance matrices in Scenario  $\Bb\mathbf{i}$. Besides, {for Scenarios  $\Cb\mathbf{2}$, we set $N=T=100,\ldots,300$ while} for Scenarios  $\Cb\mathbf{3}$-$\Cb\mathbf{5}$, we set $N=T=150$. Unsurprisingly, ``MKER" and ``MKTCR" perform more robustly. We only display the results of Scenario $\Cb\mathbf{1}$, $\Cb\mathbf{4}$, and $\Cb\mathbf{5}$ in Table \ref{tab:3}, Table \ref{tab:4} and Table \ref{tab:5} while the remaining results can be found in the Appendix.

\begin{table*}[hbpt]
  	\begin{center}
  		\addtolength{\tabcolsep}{3pt}
  		\small \caption{Simulation results for Scenario $\Cb\mathbf{1}$: $r=3,\theta=1,\rho=0.5,\beta=0.2,k_{{\rm max}}=8,J={\rm max}\{10, N/20\}$, $(\bF_t^\top,\bv_t^\top)\sim t_3(\bf{0},\bI_{N+r})$. Effects of serial and cross-sectional correlations.}\label{tab:3}
  \renewcommand{\arraystretch}{1.55}
  		\vspace{0.2cm}
  		\begin{tabular*}{16cm}{llllllll}
  			\toprule[1.2pt]
       		$N$&$T$&$r$&$\hat r_{GR}$&$\hat r_{ER}$&$\hat r_{MKER}$&$\hat r_{TCR}$&$\hat r_{MKTCR}$\\\hline
       	    25&25&3&3.194(180$|$349)&2.681(333$|$182)&3.203(125$|$295)&3.493(125$|$447)&3.728(50$|$489)\\
      	    50&50&3&3.662(85$|$470)&2.982(210$|$235)&3.634(17$|$401)&4.034(46$|$583)&4.256(2$|$646)\\
       	    75&75&3&3.436(62$|$265)&2.896(154$|$113)&3.218(10$|$113)&3.742(34$|$343)&3.869(3$|$302)\\
       	    100&100&3&3.097(52$|$122)&2.853(116$|$49)&3.013(1$|$13)&3.216(29$|$160)&3.078(0$|$38)\\
       	    125&125&3&3.069(36$|$113)&2.905(82$|$53)&3.002(0$|$2)&3.126(20$|$136)&3.003(0$|$3)\\
       	    150&150&3&3.037(25$|$78)&2.942(58$|$50)&3.000(0$|$0)&3.105(9$|$102)&3.000(0$|$0)\\
       	    175&175&3&3.048(19$|$73)&2.936(51$|$32)&3.000(0$|$0)&3.109(5$|$98)&3.000(0$|$0)\\
       	    200&200&3&3.042(12$|$63)&2.993(28$|$45)&3.000(0$|$0)&3.092(3$|$88)&3.000(0$|$0)\\
  	\bottomrule[1.2pt]		
  	\end{tabular*}\\
    \renewcommand{\thefootnote}{\fnsymbol{footnote}}
  \end{center}
\end{table*}

\begin{table*}[hbpt]
  	\begin{center}
  		\addtolength{\tabcolsep}{8.5pt}
  		\small 	\caption{Simulation results for Scenario $\Cb\mathbf{4}$: $r=3,\theta=1,\rho=0.5,\beta=0.2,J={\rm max}\{10, N/20\},N=T=150$, $(\bF_t^\top,\bv_t^\top)\sim t_3(\bf{0},\bI_{N+r})$. Effects of the choice of $k_{{\rm max}}$. }\label{tab:4}
  \renewcommand{\arraystretch}{1.55}
  		\begin{tabular*}{16cm}{lllllll}
  			\toprule[1.2pt]
       		$k_{{\rm max}}$&$r$&$\hat r_{GR}$&$\hat r_{ER}$&$\hat r_{MKER}$&$\hat r_{TCR}$&$\hat r_{MKTCR}$\\\hline
       	    8&3&3.081(19$|$101)&2.964(52$|$58)&3.000(0$|$0)&3.122(8$|$119)&3.001(0$|$1)\\
      	    12&3&3.052(23$|$87)&2.935(58$|$48)&3.000(0$|$0)&3.107(13$|$113)&3.001(0$|$1)\\
       	    16&3&3.054(24$|$85)&2.979(41$|$53)&3.000(0$|$0)&3.118(12$|$108)&3.001(0$|$1)\\
       	    20&3&3.029(24$|$72)&2.962(45$|$48)&3.000(0$|$0)&3.089(15$|$89)&3.000(0$|$0)\\
       	    25&3&3.068(24$|$98)&2.943(58$|$51)&3.000(0$|$0)&3.148(6$|$130)&3.000(0$|$0)\\
       	    30&3&3.084(21$|$101)&2.966(51$|$54)&3.001(0$|$1)&3.165(7$|$132)&3.001(0$|$1)\\
  	\bottomrule[1.2pt]		
  	\end{tabular*}\\
  \end{center}
\end{table*}

	From  Table \ref{tab:3}, we can see the proposed $\hat r_{MKER}$ and $\hat r_{MKTCR}$ still converge very quickly and show nearly exact estimation when $N,T\ge 100$, while there are almost over 10\% mis-estimation for the conventional methods even if $N,T=200$. Table \ref{tab:3} also illustrates when $N,T$ are small, $\hat r_{MKER}$ performs better than $\hat r_{MKTCR}$. From Table \ref{tab:4}, we can conclude that $k_{{\rm max}}$ still has negligible effects on the estimators even with $t_3$ samples. From Table \ref{tab:5} and the additional simulation results in the Appendix,  we conclude that ``MKTCR" tends to perform best when strong factors, weak factors or dominant factors exist. We claim that ``MKER" and ``MKTCR" always show better performances with $t_3$ samples than their corresponding competitors ``ER" and ``TCR".

    We conclude from the simulation results that the proposed two estimators perform similarly to $\hat r_{ER}$ and $\hat r_{TCR}$ when data are generated from normal distribution, and show much more stable and reliable performance when we generate data from some heavy-tailed families. The choice of $k_{{\rm max}}$ has almost no effects on the estimators. The method  $\hat r_{MKTCR}$ works well even with the existence of strong or weak factors, similar as $\hat r_{TCR}$, but more precise under the heavy-tailed cases. The different performances of these methods can also help us better understand and model real data. For example, if $\hat r_{ER}<\hat r_{GR}<\hat r_{TCR}$, there may be some strong or weak factors. Meanwhile, if $\hat r_{MKER}$ and $\hat r_{MKTCR}$ give quite different estimations compared with $\hat r_{ER}$ and $\hat r_{TCR}$, we tend to believe that the data are from  heavy-tailed distributions rather than  Gaussian distribution. It' s  further shown  in the following real data analysis {section}.

\begin{table*}[hbpt]
  	\begin{center}
  		\addtolength{\tabcolsep}{7.5pt}
  		\small 	
  		\caption{Simulation results for  Scenario   $\Cb\mathbf{5}$: $r=2,\theta=1,\rho=0.5,\beta=0.2,k_{{\rm max}}=8,J={\rm max}\{10, N/20\},N=T=150$, $(\bF_t^\top,\bv_t^\top)\sim t_3(\bf{0},\Db)$, $\Db$ is $(N+r)\times(N+r)$ diagonal with $\mathrm{D}_{ii}=1,i\ne 1;\mathrm{D}_{11}=SNR$, $SNR$ from 1 to 20. Effects of dominant factor with two factors.}\label{tab:5}
  \renewcommand{\arraystretch}{1.55}
  		\begin{tabular*}{16cm}{lllllll}
  			\toprule[1.2pt]
       		$SNR$&$r$&$\hat r_{GR}$&$\hat r_{ER}$&$\hat r_{MKER}$&$\hat r_{TCR}$&$\hat r_{MKTCR}$\\\hline
       	    1&2&2.102(14$|$109)&2.021(49$|$69)&2.000(0$|$0)&2.156(4$|$135)&2.000(0$|$0)\\
      	    3&2&2.035(36$|$67)&1.867(159$|$26)&1.996(4$|$0)&2.089(8$|$90)&2.000(0$|$0)\\
       	    7&2&1.909(135$|$44)&1.452(557$|$9)&1.902(98$|$0)&2.064(36$|$90)&1.998(2$|$0)\\
       	    10&2&1.832(205$|$33)&1.226(779$|$5)&1.736(264$|$0)&2.050(46$|$86)&2.001(0$|$1)\\
       	    15&2&1.738(282$|$20)&1.117(884$|$1)&1.513(487$|$0)&2.043(38$|$77)&1.996(5$|$1)\\
       	    20&2&1.625(389$|$13)&1.036(965$|$1)&1.257(743$|$0)&2.052(40$|$91)&1.991(9$|$0)\\
  	\bottomrule[1.2pt]		
  	\end{tabular*}\\
  \end{center}
\end{table*}

	\section{Real-data Example}\label{sec:rde}
	In this section we apply our method to a real data set FRED-MD, which was ever studied in  \cite{Xia2017Transformed}. It was collected and introduced in \cite{McCracken2015FRED}, and can be freely downloaded from the website \url{http://research.stlouisfed.org/econ/mccracken/fred-md/}. The dataset we use contains 128 monthly series of macroeconomic variables with 710 observations from 1959-01 to 2018-02. The raw data are non-stationary with missing entries. In the first step, we transform the series to stationary form with the MATLAB codes provided by the website. After this preprocessing procedure, the first two observations vanish with the difference operators and a $708\times 128$ panel remains.  The website also provide codes for replacing outliers with some ``reasonable" values, but we skip this step because extreme observations are inevitable if data are from some quite heavy-tailed distributions. Then for the missing entries in column $i$, we simply replace them with sample mean of non-missing observations in this column.
	
	We first set $k_{{\rm max}}=8$, use the whole panel and try different criteria to determine the number of common factors. To our surprise, the methods give different estimations that $\hat r_{GR}=2,\hat r_{ER}=2,\hat r_{MKER}=1,\hat r_{TCR}=5,\hat r_{MKTCR}=4$. We also tried  $k_{{\rm max}}=10,15,20,30$ and obtain completely the same results. Because $\hat r_{ER}$ gives smaller estimate than $\hat r_{TCR}$, we guess there may be some strong or weak factors. {Besides, the different estimates $\hat r_{TCR}$ and $\hat r_{MKTCR}$ imply the distribution of macroeconomic variables may be heavy-tailed, thus manipulation on outliers may be inappropriate since it brings distortion to the underlying distributions.} Overall, taking $r=4$  shall be a proper choice for the number of common factors.
	
		\begin{figure}[hbpt]
		\centering
		\includegraphics[scale=0.3]{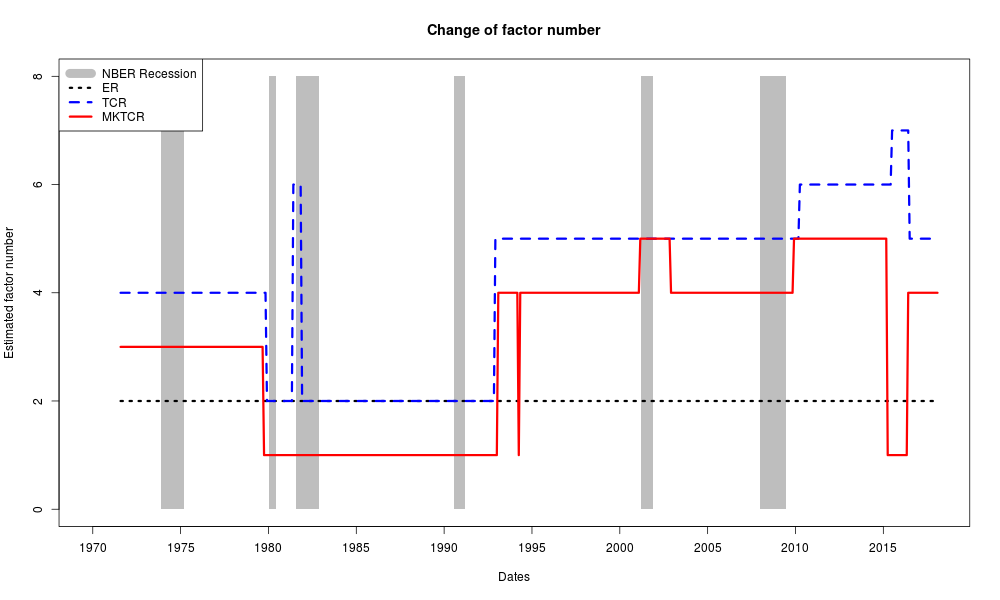}
		\caption{The change of estimated number of common factors by ``ER" (black dotted line), ``TCR" (blue dotted line), ``MKTCR" (red real line). The grey bars denote the recession dates recorded by the National Bureau of Economic Research.}
		\label{fig:3}
		\end{figure}	
	We are also interested in how the factor number changed with time. To this end, at each time  point $t$, we  repeatedly estimate the number of factors using 150 observations before (including) $t$. This is reasonable because only the past information are available in the real case. The sample size 150 is selected based on the convergence rate in simulations, then the estimated factor number series starts in 1971-08 and ends in 2018-02. Figure \ref{fig:3} shows the factor number series estimated by ``ER", ``TCR" and ``MKTCR", together with the grey bars which are the recession dates in business cycles recorded by the National Bureau of Economic Research (NBER). The NBER recession dates are open resources and available in the official website \url{http://www.nber.org/cycles.html}.

	We find that $\hat r_{ER}$ performs stably, always giving an estimate of 2 factors. In most period $\hat r_{MKTCR}$ are larger than $\hat r_{EG}$ and smaller than $\hat r_{TCR}$, which matches the simulation results when the data are generated from some heavy-tailed distributions. The kurtosis in Figure \ref{fig:1} can also support the conjecture that the real data are heavy-tailed. Another interesting finding is that both $\hat r_{TCR}$ and $\hat r_{MKTCR}$ experience several variations during or near the economic recession dates. It's possibly the recessions that cause the alternations of the number of factors. Maybe this can be applied to predict the beginning of recessions. We see that both $\hat r_{TCR}$ and $\hat r_{MKTCR}$ show a vibration from 2015 to 2016. Though it has not been recorded as recession by NEBR, it's widely known that the global economy faced with serious crisis during this period.
	
	The results of the estimated number of factors are quite different from the findings in \cite{McCracken2015FRED} in the following two aspects. Firstly McCracken and Ng \cite{McCracken2015FRED}'s method always gives larger estimates with 6 or even 8 factors.  Secondly they found that the recessions tend to increase the number of factors while in Figure \ref{fig:3} we can see that the recessions may also decrease it. McCracken and Ng \cite{McCracken2015FRED} determined the number of factors with the criterion proposed by \cite{Bai2002Determining}, and the simulations in \cite{Xia2017Transformed} give clues that  Bai and Ng \cite{Bai2002Determining}'s criteria might perform badly with serially and cross-sectionally correlated errors. From this perspective, we believe $\hat r_{MKTCR}$ and $\hat r_{TCR}$ give more reliable estimates and thus are more suitable for financial data analysis. Besides, based on our simulation results, $\hat r_{TCR}$ tends to overestimate for heavy-tailed distributed data, so we believe $\hat r_{MKTCR}$ shall be the best estimate in this real data example.
	
	\section{Conclusions}\label{sec:con}
	 We propose two estimators, named $\hat r_{MKER}$ and $\hat r_{MKTCR}$, to determine the number of common factors for heavy-tailed data. By replacing sample covariance matrix with sample multivariate Kendall's tau matrix, the new criteria remain efficient even with $t_3$, $t_2$ or Cauchy samples, compared with $\hat r_{ER}$, $\hat r_{GR}$ in \cite{Ahn2013Eigenvalue} or $\hat r_{TCR}$ in \cite{Xia2017Transformed}. When both $N,T$ go to infinity, the consistency of the proposed methods is  proved under some mild conditions. Simulation results show that the new methods perform comparably to those conventional estimators under Gaussian cases and  are significantly better  when data are generated from heavy-tailed $t$ distributions. The FRED-MD data set introduced in \cite{McCracken2015FRED} is analyzed with both new estimators and conventional ones, which provides some new perspectives and illustrates the advantage of $\hat r_{MKTCR}$.
	
	 The simulation study shows that when the errors are serially correlated, the new estimators still perform quite well. Thus we aim to relax the conditions in Assumption \ref{asum:1} to a stationary time series structure {or maybe even to the dynamic structure} in our future research. We claim that the independent assumption is only for obtaining the convergence of sample multivariate Kendall's tau matrix, where permutation and splitting techniques are involved.  The main difficulty for this extension is that it's difficult to define the stationarity for elliptical time series. The traditional weak stationarity doesn't hold because we do not assume finite moments, while the strict stationarity doesn't hold because the elliptical distributions are not closed under independent sums. On the other hand, it's also possible to extend the elliptical assumption to skewed elliptical distributions or to the cases when $\bF_t$ and $\bu_t$ are independent. We leave these extensions for future research.
	
	 Recently, Fan et al. \cite{fan2018farmtest} proposed robust covariance estimator with bounded fourth moment constraint, which is
\begin{displaymath}
\hat\bSigma_U(\tau)=\frac{2}{T(T-1)}\sum_{1\le s<t\le T}\psi_\tau\bigg(\frac{1}{2}\|\by_s-\by_t\|^2\bigg)\frac{(\by_s-\by_t)(\by_s-\by_t)^\top}{\|\by_s-\by_t\|^2},
\end{displaymath}
 where $\psi_\tau(x)={\rm min}(|x|,\tau)\text{sign}(x)$ for $x\in \mathbb{R}$. An interesting finding is that this estimator can be regarded as a combination of multivariate Kendall's tau and traditional sample covariance matrix. To see this, if we take $\tau$ as relatively small, it's almost equal to a scaled multivariate Kendall's tau matrix, while $\tau\rightarrow\infty$ corresponds to sample covariance matrix. It's possible to construct eigenvalue ratio-type estimators based on $\bSigma_U(\tau)$, which is adaptive to the tail of data distribution. However, the theoretical properties of this new approach remain unknown and the tuning for $\tau$ with cross-validation is time-consuming for practical implementation. {On the other hand, the Spearman's rho and marginal Kendall's tau correlation matrices are also widely-used to robustly estimate correlation structures. It's  a natural idea to estimate the number of common factors based on their sample versions. However, it still remains unknown whether the methods work for the high-dimensional approximate EFM framework.} We are also interested in building the theoretical framework for these new methods and providing more efficient algorithms for practical implementation in our future work.

\section*{Acknowledgements}
Yong He's research is partially supported by the grant of the National Science Foundation of China (NSFC 11801316), Natural Science Foundation
of Shandong Province (ZR2019QA002) and National Statistical Scientific Research Project (2018LY63).    Xinsheng Zhang's research is partially supported by the grant of the National Science Foundation of China (NSFC 11571080). Long Yu's research is partially supported by China Scholarship Council (No.201806100081).  

\bibliographystyle{plain}
\bibliography{RefDatBas_Main}

\clearpage
	
	\begin{center}
		APPENDIX: PROOFS OF MAIN RESULTS AND ADDITIONAL SIMULATION RESULTS
	\end{center}
	\begin{appendices}

\section{Proof of Main Results}

\noindent \textbf{Proof of Lemma \ref{lem1}.} \\
	
	Assume $\bSigma_y=\bLambda \bLambda^\top+\bSigma_u$ has the eigenvalue decomposition form $\bSigma_y=\bOmega \Hb \bOmega^\top$, with $\bOmega=(\bomega_1,\ldots,\bomega_N)$ composed of the orthogonal eigenvectors, $\Hb$ are diagonal with elements  of the ordered (decreasing) eigenvalues. Denote $\Mb=\bOmega^\top\Kb_y\bOmega$, then by the proof of Theorem 3.1 in \cite{Han2017ECA}, $\Mb$ is diagonal and
    \begin{displaymath}
    \mathrm{M}_{jj}={\rm E}\left(\frac{\mathrm{H}_{jj}U_j^2}{\mathrm{H}_{11}U_1^2+\cdots+\mathrm{H}_{q^\star q^\star}U_{q^\star}^2}\right), j=1,\ldots,q^\star,
    \end{displaymath}	
    where $\bU=(U_1,\ldots,U_{q^\star})$ are uniform random vector from the unit sphere in $\mathbb{R}^{q^\star}$, $q^\star=\text{rank}(\bSigma_y)$. Han  and Liu \cite{Han2017ECA} also proved $\mathrm{M}_{jj}\ge \mathrm{M}_{j+1,j+1}$ for $j=1,\ldots,q^\star-1$, so $\mathrm{M}_{jj}=\lambda_j(\Kb_y)$. Firstly, we will give the upper and lowers bounds for $\mathrm{M}_{jj}$.

    By Assumption \ref{asum:3} and Weyl's theorem, $\lambda_j(\bLambda^\top\bLambda/N)=\lambda_j(\bSigma_{\bLambda})+o(1)$ for $j=1,\ldots,r$. Thus the eigenvalues of $\bLambda\bLambda^\top$ are
    \begin{displaymath}
    N\{\lambda_1(\bSigma_{\bLambda})+o(1)\},\ldots,N\{\lambda_r(\bSigma_{\bLambda})+o(1)\},0,\ldots,0.
    \end{displaymath}

   By Assumption \ref{asum:2} and Weyl's theorem again, we have $N\{C_2+o(1)\}\le \mathrm{H}_{jj}\le N\{C_1+o(1)\}$ for $j\le r$, $C_2\le \mathrm{H}_{jj}\le C_1$ for  $r<j\le m$ and $\mathrm{H}_{jj}\le C_1$ for $m<j\le N$, with $m={\rm min}\{N,T\}$. Thus we have $q^\star\ge m$.

   It's easy to check for $j\le r$ and $N\rightarrow \infty$,
	\begin{displaymath}
	 \mathrm{M}_{jj}\le\frac{\mathrm{H}_{11}}{\mathrm{H}_{rr}}{\rm E}\left(\frac{U_j^2}{U_1^2+\cdots+U_r^2}\right)\le\frac{C_1}{rC_2}+o(1).
	\end{displaymath}
	For the lower bound, note that	
	\begin{displaymath}
	\begin{split}
	\mathrm{M}_{jj}\ge&{\rm E}\left(\frac{\mathrm{H}_{rr}U_j^2}{\mathrm{H}_{11}U_1^2+\cdots+\mathrm{H}_{11}U_r^2+C_1U_{r+1}^2+\cdots+C_1U_{q^\star}^2}\right)\\
	=&{\rm E}\left\{\frac{\mathrm{H}_{rr}U_j^2}{(\mathrm{H}_{11}-C_1)U_1^2+\cdots+(\mathrm{H}_{11}-C_1)U_r^2+C_1}\right\}\\
	=&\frac{\mathrm{H}_{rr}}{(\mathrm{H}_{11}-C_1)}{\rm E}\left\{\frac{(\mathrm{H}_{11}/C_1-1)U_j^2}{(\mathrm{H}_{11}/C_1-1)U_1^2
+\cdots+(\mathrm{H}_{11}/C_1-1)U_r^2+1}\right\}.
	\end{split}
	\end{displaymath}
	For $j\le r$, define \[
a_N=\mathrm{H}_{11}/C_1-1,\quad \text{and} \quad b_N={\rm E}\left(\frac{a_NU_1^2}{1+a_NU_1^2+\cdots+a_NU_r^2}\right)={\rm E}\left(\frac{a_NU_j^2}{1+a_NU_1^2+\cdots+a_NU_r^2}\right),
\]  then we have
	\begin{displaymath}
	rb_N+{\rm E}\left(\frac{1}{1+a_NU_1^2+\cdots+a_NU_r^2}\right)=1,
	\end{displaymath}
	and
	\begin{displaymath}
	{\rm E}\left(\frac{1}{1+a_NU_1^2+\cdots+a_NU_r^2}\right)\le{\rm E}\bigg(\frac{1}{1+a_NU_1^2}\bigg).
	\end{displaymath}\\
	Let $\mathcal{T}=I\{a_NU_1^2>1\}$, then with $N$ sufficiently large we have $a_N>0$ and $a_NU_1^2>\mathcal{T}$. Thus
 \[
{\rm E}\left(\frac{1}{1+a_NU_1^2}\right)\le{\rm E}\left(\frac{1}{1+\mathcal{T}}\right)=\Pr(\mathcal{T}=0)+\frac{1}{2}\Pr(\mathcal{T}=1)
=\frac{1}{2}\Pr(\mathcal{T}=0)+\frac{1}{2}.
\]
Note that $U_1\stackrel{d}{=}{g_1}/{\|\bg\|}$, where $\bg=(g_1,\ldots,g_{q^\star})\sim \mathcal{N}({\bf 0},\Ib_{q^\star})$. Then, with $m\rightarrow\infty$,
	\begin{displaymath}
	\Pr(\mathcal{T}=0)=\Pr\bigg(g_1^2<\frac{1}{a_N-1}\sum_{i=2}^{q^\star}g_i^2\bigg)\le \Pr(g_1^2<{q^\star}^{-1/4})+\Pr\bigg(\frac{1}{a_N-1}\sum_{i=2}^{q^\star}g_i^2\ge {q^\star}^{-1/4}\bigg)\rightarrow 0.
	\end{displaymath}
	Then, we have
	\begin{displaymath}
	{\rm E}\left(\frac{1}{1+a_NU_1^2}\right)\le \frac{1}{2}+o(1) \quad \text{and} \quad b_N\ge\frac{1}{r}(\frac{1}{2}+o(1))\Rightarrow \mathrm{M}_{jj}\ge \frac{C_2}{2rC_1}+o(1),j\le r.
	\end{displaymath}
	For $r<j\le m$, with $m={\rm min}\{N,T\}$, we have
\[
\mathrm{M}_{jj}\le{\rm E}\left(\frac{C_1U_j^2}{C_2U_{r+1}^2+\cdots+C_2U_m^2}\right)
=\frac{C_1}{(m-r)C_2}=\frac{C_1}{mC_2}\{1+o(1)\}=O(m^{-1}).
\]
For $j>m$, $\mathrm{M}_{jj}\le\mathrm{M}_{mm}=O({1}/{m})$. We don't need the lower bound of $\mathrm{M}_{jj}$ for ${j>r}$.\qed\\

To prove Theorem \ref{the1}, we need the next Lemma \ref{lemma:a1} to bound the asymptotic distance between $\hat{\Kb}_y$ and $\Kb_y$. The lemma was adapted from the proof of Theorem 4.1, Lemmas F.1 and F.2 in \cite{fan2018}. We point out that Fan et al. \cite{fan2018} assumed bounded fourth moment only for the sake of estimating marginal variances by M-estimators with Huber loss,  as what they claimed in \cite{fan2018}. We claim that the finite fourth moment condition is not needed in the proof of our Lemma \ref{lemma:a1}. We organized the proof of  our Lemma \ref{lemma:a1} into 4 steps to carefully address the matter.
 \begin{lemma}\label{lemma:a1}
Assume Assumptions \ref{asum:1}-\ref{asum:3} hold, we have $\|\hat \Kb_y-\Kb_y\|=O_p(rm^{-1/2})$, with $m={\rm min}\{N,T\}$.
 \end{lemma}
Proof. Here is an outline for the proof. \begin{description}
	\item[Step 1] Assume $T$ is even with $\bar t=T/2$, and define $\tilde w_s=\Hb^{\frac{1}{2}}\bg_s/(\bg_s^\top\Hb\bg_s)^{\frac{1}{2}}$ for $s=1,\ldots,\bar t$ with $\bg_s\sim\mathcal{N}_N({\bf 0},\Ib)$ are independent Gaussian vectors and $\Hb$ defined preliminarily, $\hat \Kb_g=\bar t^{-1}\sum_{s=1}^{\bar t}\tilde\bw_s\tilde\bw_s^\top$ and $\Kb_g={\rm E}(\hat\Kb_g)$. We will show in order to prove Lemma \ref{lemma:a1}, it suffices to prove ${\rm E}\|\hat\Kb_g-\Kb_g\|=O(rm^{-1/2})$.
	\item[Step 2] We will show  $|\lambda_j(\hat\Kb_g)-\lambda_j(\Kb_g)|=O_p(\sqrt{r/m})$ for $j\le r$ and $\lambda_j(\hat\Kb_g)=O_p(m^{-1})$ for $j>r$.
	\item[Step 3] Denote the eigenvector matrix of $\hat\Kb_g$ as $\hat\bGamma=(\hat\bGamma_1,\hat \bGamma_2)$ with $\hat\bGamma_1$ composed of the leading $r$ eigenvectors, we will show $\|\hat\bGamma_1-(\Ib_r,{\bf 0})^\top\|=O_p(r^{2}m^{-1/2})$.
	\item[Step 4] Combining the above results to conclude our lemma.
\end{description}
Now we move to the detailed proofs for each step.\\

\textbf{Proof of Step 1}.
Note that by Assumption \ref{asum:1}, $\by_t\sim EC_N(\bmu_y,\bSigma_y,\xi)$, where $\bmu_y=(\bLambda,\Ib_N)\bmu_0$ and $\bSigma_y=\bLambda\bLambda^\top+\bSigma_u=\bOmega\Hb\bOmega^\top.$ Define $\bz_t=\bOmega^\top\by_t,t=1,\ldots,T$, then $\bz_t\sim EC_N(\bmu_z,\Hb,\xi)$, where $\bmu_z=\bOmega^\top\bmu_y$. Construct the sample and population multivariate Kendall's matrices with $\bz_t$ by
	\begin{displaymath}
	\Kb_z={\rm E}\bigg\{\frac{(\bz_1-\bz_2)(\bz_1-\bz_2)^\top}{\|\bz_1-\bz_2\|^2}\bigg\},\quad\hat \Kb_z=\frac{2}{T(T-1)}\sum\limits_{1\le i<j\le T}\frac{(\bz_i-\bz_j)(\bz_i-\bz_j)^\top}{\|\bz_i-\bz_j\|^2},
	\end{displaymath}
 then $\Kb_z=\bOmega^\top\Kb_y\bOmega=\Mb$, $\hat\Kb_z=\bOmega^\top\hat\Kb_y\bOmega$ and
 \begin{displaymath}
 \|\hat \Kb_y-\Kb_y\|=\|\bOmega(\hat \Kb_z-\Kb_z)\bOmega^\top\|=\|\hat \Kb_z-\Kb_z\|.
 \end{displaymath}

 Assume $T$ is even and $\bar t=T/2$, otherwise we can delete the last observation. For any permutation $\sigma$ of $\{1,\ldots,T\}$, define $\bz_t^\sigma$ as the corresponding $t$-th observation after permutation. Define $\bw_s^\sigma=(\bz_{2s-1}^\sigma-\bz_{2s}^\sigma)/\|\bz_{2s-1}^\sigma-\bz_{2s}^\sigma\|$ for $ s=1,\ldots,\bar t$, and $\hat\Kb_z^\sigma={\bar t}^{-1}\sum_{s=1}^{\bar t}\bw_s^\sigma{\bw_s^\sigma}^\top$, then
 \begin{displaymath}
 \sum_{\sigma\in\mathcal{S}_T}\bar t\hat\Kb_z^\sigma=T\times(T-2)!\times\frac{T(T-1)}{2}\hat\Kb_z\Rightarrow \hat\Kb_z=\frac{1}{T!}\sum_{\sigma\in\mathcal{S}_T}\hat\Kb_z^\sigma,
 \end{displaymath}
 where $\mathcal{S}_T$ is the permutation group of $\{1,\ldots,T\}$. So,
 \begin{displaymath}
\|\hat \Kb_z-\Kb_z\|=\|\frac{1}{T!}\sum_{\sigma\in\mathcal{S}_T}(\hat\Kb_z^\sigma-\Kb_z)\|\le\frac{1}{\text{card}(\mathcal{S}_T)}\sum_{\sigma\in\mathcal{S}_T}\|(\hat\Kb_z^\sigma-\Kb_z)\|,
 \end{displaymath}
 and it suffices to show ${\rm E}\|(\hat\Kb_z^\sigma-\Kb_z)\|=O(r/\sqrt{m})$ for any given $\sigma$.\\

 We regard $\sigma$ as given and define $\tilde\bw_s=\Hb^{\frac{1}{2}}\bg_s/(\bg_s^\top\Hb\bg_s)^{\frac{1}{2}}$ for $s=1,\ldots,\bar t$, with some independent Gaussian vectors $\bg_s=(g_{s1},\ldots,g_{sN})\sim\mathcal{N}(\bf0,\Ib_N)$. It's easy to see $\hat\bw_s^\sigma\stackrel{d}{=}\tilde\bw_s$ because $(\bz_{2s-1}^\sigma-\bz_{2s}^\sigma)\stackrel{d}{=}\Hb^{\frac{1}{2}}\bg_s/\|\bg_s\|$ by the stochastic representation of elliptical vectors. Therefore,  $\hat \bw_s^\sigma$ and $\tilde \bw_s$ can be regarded as sampled from the same distribution. Define $\hat\Kb_g={\bar t}^{-1}\sum_{s=1}^{\bar t}\tilde\bw_s{\tilde\bw_s}^\top$ and $\Kb_g={\rm E}(\hat\Kb_g)={\rm E}(\tilde\bw_1{\tilde\bw_1}^\top)$, then $\Kb_g=\Kb_z$ and $\|\hat\Kb_g-\Kb_g\|$ shares the same asymptotic properties as  $\|\hat\Kb_z^\sigma-\Kb_z\|$. As a result, it suffices to show ${\rm E}\|\hat\Kb_g-\Kb_g\|=O(rm^{-1/2})$.

 	\textbf{Step 1} makes it clear that the consistency of $\Kb_y$ doesn't depend on the constraint of bounded fourth moments. Only $\Hb$, which is closely related to the scatter matrix $\bSigma_y$, determines the consistent rate.  \qed\\

 \textbf{Proof of Step 2}. We need more notations on this step.  Define
 \begin{displaymath}
\Lb_{\bar t\times \bar t}={\rm diag}\big\{(N^{-1}\bg_1^\top\Hb\bg_1)^{-\frac{1}{2}},\ldots,(N^{-1}\bg_{\bar t}^\top\Hb\bg_{\bar t})^{-\frac{1}{2}}\big\},
 \end{displaymath}
 $\bm\eta_j=\Lb(g_{1j},\ldots,g_{\bar tj})^\top/\sqrt{N}$, and $\tilde \Wb=(\tilde\bw_1,\ldots,\tilde\bw_{\bar t})^\top=(\bm\eta_1,\ldots,\bm\eta_{N})\Hb^{\frac{1}{2}}$, then $\hat\Kb_g=\bar t^{-1}\tilde{\Wb}^\top\tilde \Wb$.  Let $\tilde\Kb_g=\bar t^{-1}\tilde{\Wb}\tilde \Wb^\top$, then $\hat\Kb_g$ and $\tilde\Kb_g$ share the same non-zero eigenvalues.
 Separate $\tilde\Wb$  as $\tilde\Wb=(\Ab,\Bb)\Hb^{1/2}=(\Ab\Hb_A^{1/2},\Bb\Hb_B^{1/2})$ with $\Ab=(\bm\eta_1,\ldots,\bm\eta_r)$,  $\Bb=(\bm\eta_{r+1},\ldots,\bm\eta_N)$, $\Hb_{A}={\rm diag}(\mathrm{H}_{11},\ldots,\mathrm{H}_{rr})$ and $\Hb_{B}={\rm diag}(\mathrm{H}_{r+1,r+1},$ $\ldots,\mathrm{H}_{NN})$. Obviously $\tilde\Kb_g=\bar t^{-1}(\Ab\Hb_A\Ab^\top+\Bb\Hb_B\Bb^\top)$. It suffices to  bound the two parts of $\tilde\Kb_g$.

We first deal with $\mathcal{A}=\bar t^{-1}\Ab\Hb_A\Ab^\top$, whose rank is at most $r$. The non-zero eigenvalues of $\mathcal{A}$ are equal to those of $\tilde{\mathcal{A}}=\bar t^{-1}\Hb_A^{1/2}\Ab^\top\Ab\Hb_A^{1/2}$. $\tilde{\mathcal{A}}$ is $r\times r$ symmetric matrix with $ij$-th entry as
\begin{displaymath}
\tilde{\mathcal{A}}_{ij}=\bar t^{-1}(\mathrm{H}_{ii}\mathrm{H}_{jj})^{1/2}\sum_{s=1}^{\bar t}\frac{g_{si}g_{sj}}{\bg_s^\top\Hb\bg_s}.
\end{displaymath}
By central limit theorem, $\tilde{\mathcal{A}}_{jj}=\mathrm{M}_{jj}+O_p(\bar t^{-1/2})$ for $j\le r$, while $\mathrm{M}_{jj}=\lambda_j(\Kb_z)=\lambda_j(\Kb_g)$. Also, $\tilde{\mathcal{A}}_{ij}=O_p(1/\sqrt{r\bar t})$ for $i<j\le r$. Therefore, $\lambda_j(\mathcal{A})=\lambda_j(\Kb_g)+O_p(\sqrt{r/\bar t})$ for $j\le r$, while $\lambda_j(\mathcal{A})=0$ for $j>r$.

Next, define $\mathcal{B}=\bar t^{-1}\Bb\Hb_B\Bb^\top$ and $\Rb$ is $\bar t\times(N-r)$ matrix with $\Rb_{sj}=g_{s,r+j}/\sqrt{N}$ for $s\le \bar t$ and $j\le N-r$, thus $\Bb=\Lb\Rb$, and
\begin{displaymath}
\bar t\|\mathcal{B}\|=\|\Lb\Rb\Hb_B\Rb^\top\Lb^\top\|\le \|\Lb\|^2\|\Rb\Hb_B\Rb^\top\|.
\end{displaymath}
Since the rows of $\Rb$ are i.i.d. Gaussian vectors, by Lemma D.1 in \cite{fan2018},
\begin{displaymath}
\|\Rb\Hb_B\Rb^\top\|=\frac{1}{N-r}\sum_{j=r+1}^{N}\mathrm{H}_{jj}+O_p\bigg(\frac{T}{N}+\sqrt{\frac{T}{N}}\bigg).
\end{displaymath}
And $\|\Lb\|^2={\rm max}_{s\le \bar t}(N^{-1}\bg_s^\top\Hb\bg_s)^{-1}\le C_2^{-1}+o_p(1)$. Then, $\|\mathcal{B}\|=O_p(m^{-1})$.

Combine the above results and Weyl's theorem, $|\lambda_j(\hat\Kb_g)-\lambda_j(\Kb_g)|=O_p(\sqrt{r/m})$ for $j\le r$ and $\lambda_j(\hat \Kb_g)=O_p(m^{-1})$ for $j>r$.\qed\\

\textbf{Proof of Step 3}. Denote $\hat\bGamma_1=\{\gamma_{ij}\}$ with $i\le N$ and $j\le r$.  We first show that $\sum_{i=r+1}^{N}\gamma_{ij}^2=O_p(rm^{-1})$ for any $j\le r$.  Define a $N\times N$ diagonal matrix $\tilde\Hb$ by $\mathrm{\tilde H}_{ii}=N^{-1}$ for $i\le r$ and $\mathrm{\tilde H}_{ii}=1$ otherwise.  Given $j\le r$, we then have
\begin{displaymath}
\sum_{i=1}^{N}\mathrm{\tilde H}_{ii}\gamma_{ij}^2=\bgamma_j^\top\tilde\Hb\bgamma_j=\|\tilde\Hb^{\frac{1}{2}}\bgamma_j\bgamma_j^\top\tilde\Hb^{\frac{1}{2}}\|\le M_{jj}^{-1}\|\tilde\Hb^{\frac{1}{2}}\hat\Kb_g\tilde\Hb^{\frac{1}{2}}\|=M_{jj}^{-1}\|\bar t^{-1}(\Ab,\Bb)\mathcal{H}(\Ab,\Bb)^\top\|,
\end{displaymath}
where $\mathcal{H}=\Hb^{1/2}\tilde\Hb\Hb^{1/2}$. For any $i\le N$, $\mathcal{H}_{ii}\asymp 1$. By similar technique as when we bound $\|\mathcal{B}\|$, with $\mathrm{M_{jj}}=O_p(r^{-1})$, we can get
\begin{displaymath}
\sum_{i=1}^{N}\mathrm{\tilde H}_{ii}\gamma_{ij}^2=O_p(rm^{-1}).
\end{displaymath}
Note that $\mathrm{\tilde H}_{ii}=1$ for $i>r$,  so $\sum_{i=r+1}^{N}\gamma_{ij}^2=O_p(rm^{-1})$ and $\sum_{i=1}^{r}\gamma_{ij}^2=1+O_p(rm^{-1})$ for any $j\le r$.

Next denote $\bdelta_j$ as the $j$-th eigenvector of $\tilde\Kb_g$. By some algebra,
\begin{equation}\label{equ:a1}
(\mathrm{M}_{jj}\Ib_{\bar t}-\mathcal{A})\bdelta_j=\mathcal{B}\bdelta_j-\{\lambda_j(\hat\Kb_g)-\mathrm{M}_{jj}\}\bdelta_j.
\end{equation}
Define $\bgamma_j^{(1)}=(\gamma_{1j},\ldots,\gamma_{rj})$ and $\bgamma_j^{(2)}=(\gamma_{r+1,j},\ldots,\gamma_{Nj})$, then by the relationship
\begin{displaymath}
\bgamma_j=\frac{\tilde\Wb^\top\bdelta_j}{\sqrt{\bar t\lambda_j(\hat\Kb_g)}},
\end{displaymath}
Left multiply  $\Hb_A^{1/2}\Ab^\top/\sqrt{\bar t\lambda_j(\hat\Kb_g)}$ to both sides of Equation (\ref{equ:a1}) to obtain
\begin{equation}\label{equ:a2}
\begin{split}
(\mathrm{M}_{jj}\Ib_{r}-\bTheta_A)\bgamma_j^{(1)}=&(\bar t^{-1}\Hb_A^{1/2}\Ab^\top\Ab\Hb_A^{1/2}-\bTheta_A)\bgamma_j^{(1)}+\bar t^{-1}\Hb_A^{1/2}\Ab^\top\Bb\Hb_B^{1/2}\bgamma_j^{(2)}\\
&-\bigg\{\lambda_j(\hat\Kb_g)-\mathrm{M}_{jj}\bigg\}\bgamma_j^{(1)},
\end{split}
\end{equation}
where $\bTheta_A$ is $r\times r$ diagonal with $\bTheta_{A,kk}=\mathrm{M}_{kk}$ for $k\le r$.  Next, define
\begin{displaymath}
\Qb=\sum_{k\le r,k\ne j}\frac{1}{\mathrm{M}_{jj}-\mathrm{M}_{kk}}\be_k\be_k^\top,
\end{displaymath}
where $\be_k$ is $r$-dimensional vector with $k$-th entry equal to 1 while other entries are 0. It's easy to check $\Qb(\mathrm{M}_{jj}\Ib_r-\bTheta_A)=\sum\limits_{k\ne j}\be_k\be_k^\top=\Ib_r-\be_j\be_j^\top$. Left multiply $\Qb$ to Equation (\ref{equ:a2}) and then divide $\|\bgamma_l^{(1)}\|$ to obtain
\begin{displaymath}
\begin{split}
\bv_j:=\frac{\bgamma_j^{(1)}}{\|\bgamma_j^{(1)}\|}-\langle\frac{\bgamma_j^{(1)}}{\|\bgamma_j^{(1)}},\be_j\rangle\be_j=&\Qb(\bar t^{-1}\Hb_A^{1/2}\Ab^\top\Ab\Hb_A^{1/2}-\bTheta_A)\frac{\bgamma_j^{(1)}}{\|\bgamma_j^{(1)}\|}\\
&+\bar t^{-1}\Qb\Hb_A^{1/2}\Ab^\top\Bb\Hb_B^{1/2}\frac{\bgamma_j^{(2)}}{\|\bgamma_j^{(1)}\|}
-\bigg\{\lambda_j(\hat\Kb_g)-\mathrm{M}_{jj}\bigg\}\Qb\frac{\bgamma_j^{(1)}}{\|\bgamma_j^{(1)}\|}.
\end{split}
\end{displaymath}
We shall bound the error terms one by one. Firstly for any $j<k\le r$, it's not difficult to verify
\begin{displaymath}
\mathrm{M}_{jj}-\mathrm{M}_{kk}\ge O(r^{-1}).
\end{displaymath}
Therefore, $\|\Qb\|=O_p(r)$. By step 2, $\|\bar t^{-1}\Hb_A^{1/2}\Ab^\top\Ab\Hb_A^{1/2}-\bTheta_A\|=O_p(\sqrt{r/m})$. And,
\begin{displaymath}
\bigg\|\bar t^{-1}\Qb\Hb_A^{1/2}\Ab^\top\Bb\Hb_B^{1/2}\frac{\bgamma_j^{(2)}}{\|\bgamma_j^{(1)}\|}\bigg\|\le\|\Qb\|\|\mathcal{A}\|^{1/2}\|\mathcal{B}\|^{1/2}\bigg\|\frac{\bgamma_j^{(2)}}{\|\bgamma_j^{(1)}\|}\bigg\|=O_p(\sqrt{r/m}).
\end{displaymath}
Combine the above bounds to obtain $\|\bv_j\|=O_p(r^{3/2}m^{-1/2})$.

Further, project $\bgamma_j^{(1)}/\|\bgamma_j^{(1)}\|$ to $\be_j$ so that
\begin{displaymath}
\frac{\bgamma_j^{(1)}}{\|\bgamma_j^{(1)}\|}=\bv_j+\langle\frac{\bgamma_j^{(1)}}{\|\bgamma_j^{(1)}},\be_j\rangle\be_j,
\end{displaymath}
while $\bv_j$ and $\be_j$ are orthogonal. Therefore,
\begin{displaymath}
1-\langle\frac{\bgamma_j^{(1)}}{\|\bgamma_j^{(1)}},\be_j\rangle=1-\sqrt{1-\|\bv_j\|^2}\le\|\bv_j\|^2.
\end{displaymath}
We also have
\begin{displaymath}
\frac{\bgamma_j^{(1)}}{\|\bgamma_j^{(1)}\|}-\be_j=\bv_j+\bigg(\langle\frac{\bgamma_j^{(1)}}{\|\bgamma_j^{(1)}},\be_j\rangle-1\bigg)\be_j,
\end{displaymath}
which shows
\begin{displaymath}
\bigg\|\frac{\bgamma_j^{(1)}}{\|\bgamma_j^{(1)}\|}-\be_j\bigg\|\le\|\bv_j\|+\|\bv_j\|^2=O_p(r^{3/2}m^{-1/2}).
\end{displaymath}
Combined with $\|\bgamma_j^{(1)}\|^2=1+O_p(rm^{-1})$, we have $\|\bgamma_j^{(1)}-\be_j\|=O_p(r^{3/2}m^{-1/2})$. And
\begin{displaymath}
\|\hat\bGamma_1-(\Ib_r,{\bf 0})^\top\|\le\bigg\{\sum_{j=1}^{r}(\|\bgamma_j^{(1)}-\be_j\|^2+\|\bgamma_j^{(2)}\|^2)\bigg\}^{1/2}=O_p(r^2m^{-1/2}).
\end{displaymath}\qed

\textbf{Proof of Step 4}.
Define $\bTheta_A={\rm diag}(\mathrm{M}_{11},\ldots,\mathrm{M}_{rr})$,  $\bTheta_B={\rm diag}(\mathrm{M}_{r+1,r+1},\ldots,\mathrm{M}_{NN})$, which are composed of the eigenvalues of $\Kb_g$, and $\hat\bTheta_A={\rm diag}\{\lambda_1(\hat \Kb_g),\ldots,\lambda_r(\hat \Kb_g)\}$, $\hat\bTheta_B={\rm diag}\{\lambda_{r+1}(\hat \Kb_g),\ldots,\lambda_N(\hat \Kb_g)\}$. We have
\begin{displaymath}
\begin{array}{llll}
\|\hat\Kb_g-\Kb_g\|&=&\|(\hat\bGamma_1,\hat \bGamma_2){\rm diag}(\hat\bTheta_A,\hat\bTheta_B)(\hat\bGamma_1,\hat \bGamma_2)^\top-{\rm diag}(\bTheta_A,\bTheta_B)\|\\
&\le&\|\hat\bGamma_1\hat\bTheta_A\hat\bGamma_1^\top-{\rm diag}(\bTheta_A,\bm 0)\|+\|\hat\bGamma_2\hat\bTheta_B\hat\bGamma_2^\top-{\rm diag}(\bm 0,\bTheta_B)\|\\
&\le&\|\hat\bGamma_1(\hat\bTheta_A-\bTheta_A)\hat\bGamma_1^\top\|+\|\hat\bGamma_1\bTheta_A\{\hat\bGamma_1-(\Ib_r,\bm{0})^\top\}^\top\|+\|\{\hat\bGamma_1-(\Ib_r,\bm{0})^\top\}\bTheta_A(\Ib_r,\bm{0})\|\\
&&+\|\hat\bGamma_2\|^2\|\hat\bTheta_B\|+\|\bTheta_B\|\\
&\le&\|\hat\bGamma_1\|^2\|\hat\bTheta_A-\bTheta_A\|+\|\hat\bGamma_1\|\|\bTheta_A\|\hat\bGamma_1-(\Ib_r,\bm{0})^\top\|+\|\bTheta_A\|\hat\bGamma_1-(\Ib_r,\bm{0})^\top\|\\
&&+\|\hat\bGamma_2\|^2\|\hat\bTheta_B\|+\|\bTheta_B\|\\
&=&O_p(rm^{-1/2}).
\end{array}
\end{displaymath}
By the proof of Theorem 4.1 in \cite{fan2018}, the convergence can be improved to ${\rm E}\|\hat\Kb_g-\Kb_g\|=O(rm^{-1/2})$ by detailed investigation of the proof, which concludes the result of Lemma \ref{lemma:a1}.\qed

\section*{Proof of Theorem \ref{the1}.} Denote $\delta_{NT}=m^{-1/2}$, then by Lemma \ref{lem1}, Lemma \ref{lemma:a1} as well as  Weyl's theorem,
	\begin{displaymath}
	\lambda_j(\hat \Kb_y)\asymp 1,j\le r;\quad \lambda_j(\hat \Kb_y)=O_p(\delta_{NT}),j>r.
	\end{displaymath}
	For the modified empirical eigenvalues $\hat\lambda_j(\hat \Kb_y)=\lambda_j(\hat \Kb_y)+c\delta_{NT}$, we have $\hat\lambda_j(\hat \Kb_y)\asymp 1,j\le r$ and $\hat\lambda_j(\hat \Kb_y)\asymp\delta_{NT},j>r$. Now we can show the consistency of $\hat r_{MKER}$ and $\hat r_{MKTCR}$ with $\hat\lambda_j(\hat \Kb_y)$.\\
	
	It's easy to check for $j<r$ and $j>r$, $\hat\lambda_j(\hat \Kb_y)/\hat\lambda_{j+1}(\hat \Kb_y)\asymp1$, while for $j=r$ we have $\hat\lambda_r(\hat \Kb_y)/\hat\lambda_{r+1}(\hat \Kb_y)\asymp \delta_{NT}^{-1}\rightarrow \infty$, then $\hat r_{MKER}$ is consistent. And for $\hat r_{MKTCR}$,  apply the inequality $c/(1+c)<\ln(1+c)<c$ with $c>0$, then for $j<r$ or $r<j\le m-1$,
	\begin{displaymath}
	\frac{\ln\{1+\hat\lambda_j(\hat \Kb_y)/V_{j-1}\}}{\ln\{1+\hat\lambda_{j+1}(\hat \Kb_y)/V_{j}\}}<\frac{\hat\lambda_j(\hat \Kb_y)}{V_{j-1}}\cdot\frac{1+\hat\lambda_{j+1}(\hat \Kb_y)/V_j}{\hat\lambda_{j+1}(\hat \Kb_y)/V_j}=\frac{\hat\lambda_j(\hat \Kb_y)}{\hat\lambda_{j+1}(\hat \Kb_y)}\cdot\frac{V_j+\hat\lambda_{j+1}(\hat \Kb_y)}{V_j+\hat\lambda_j(\hat \Kb_y)}\le\frac{\hat\lambda_j(\hat \Kb_y)}{\hat\lambda_{j+1}(\hat \Kb_y)}=O_p(1),
	\end{displaymath}
	where $V_j=\sum_{i=j+1}^{m}\hat\lambda_i(\hat \Kb_y),j=0,\ldots,m-1$. On the other hand, for $j=r$,
	\begin{displaymath}
	\frac{\ln\{1+\hat\lambda_r(\hat \Kb_y)/V_{r-1}\}}{\ln\{1+\hat\lambda_{r+1}(\hat \Kb_y)/V_{r}\}}>\frac{\hat\lambda_r(\hat \Kb_y)}{\hat\lambda_{r+1}(\hat \Kb_y)}\cdot\frac{V_r}{V_{r-1}+\hat\lambda_r(\hat \Kb_y)}\asymp\frac{1}{\delta_{NT}}\cdot\frac{m\delta_{NT}}{m\delta_{NT}+1} \asymp\frac{1}{\delta_{NT}+1/m}\rightarrow\infty,
	\end{displaymath}
	where $m={\rm min}\{N,T\}$, as defined in Section 3. Thus $\hat r_{MKTCR}$ is consistent.\qed	

\clearpage

\section{Additional Simulation Results}

\begin{table*}[hbpt]
  	\begin{center}
  		\addtolength{\tabcolsep}{2.5pt}
  		\small \caption{Simulation results for Scenario $\Bb\mathbf{2}$: $r=3,\theta=6,\rho=0.5,\beta=0.2,k_{{\rm max}}=8,J={\rm max}\{10, N/20\}$, $(\bF_t^\top,\bv_t^\top)\sim \mathcal{N}(\bf{0},\bI_{N+r})$.  Effects of all weak factors.}\label{tab:6}
    \renewcommand{\arraystretch}{1.55}
  		\vspace{0.2cm}
  		\begin{tabular*}{16cm}{llllllll}
  			\toprule[1.2pt]
      		$N$&$T$&$r$&$\hat r_{GR}$&$\hat r_{ER}$&$\hat r_{MKER}$&$\hat r_{TCR}$&$\hat r_{MKTCR}$\\\hline
       	   25&25&3&3.325(348$|$459)&2.798(467$|$329)&3.337(349$|$451)&3.432(328$|$485)&3.684(273$|$534)\\
      	    50&50&3&4.932(51$|$904)&4.428(133$|$801)&4.749(68$|$878)&5.047(38$|$923)&5.148(20$|$937)\\
       	    75&75&3&6.364(29$|$950)&5.783(89$|$881)&6.155(39$|$937)&6.501(21$|$964)&6.615(11$|$980)\\
       	    100&100&3&6.434(72$|$898)&5.721(153$|$794)&6.186(93$|$866)&6.569(60$|$915)&6.771(39$|$939)\\
       	    125&125&3&4.798(209$|$608)&4.122(301$|$484)&4.628(224$|$577)&4.893(198$|$624)&5.319(146$|$704)\\
       	    150&150&3&3.626(214$|$310)&3.179(291$|$212)&3.567(221$|$301)&3.690(205$|$324)&3.983(174$|$390)\\
       	    175&175&3&3.025(138$|$94)&2.871(188$|$68)&2.964(177$|$88)&3.083(128$|$110)&3.19(120$|$133)\\
       	    200&200&3&2.961(50$|$23)&2.910(81$|$16)&2.937(66$|$21)&2.981(44$|$28)&3.016(33$|$36)\\
  	\bottomrule[1.2pt]		
  	\end{tabular*}\\
  \end{center}
\end{table*}

\vspace{5em}

\begin{table*}[hbpt]
  	\begin{center}
  		\addtolength{\tabcolsep}{7pt}
  		\small \caption{Simulation results for Scenario   $\Bb\mathbf{3}$: $r=3,\theta=1,\rho=0.5,\beta=0.2,k_{{\rm max}}=8,J={\rm max}\{10, N/20\},N=T=100$, $(\bF_t^\top,\bv_t^\top)\sim \mathcal{N}(\bf{0},\Db)$, $\Db$ is $(N+r)\times(N+r)$ diagonal with $\mathrm{D}_{ii}=1,i\ne 3;\mathrm{D}_{33}=SNR$, $SNR$ from 0.7 to 0.4. Effects of strong and weak factors, Gaussian samples}
  		\vspace{0.2cm}\label{tab:7}
    \renewcommand{\arraystretch}{1.55}
  		\begin{tabular*}{16cm}{lllllll}
  			\toprule[1.2pt]
       		$SNR$&$r$&$\hat r_{GR}$&$\hat r_{ER}$&$\hat r_{MKER}$&$\hat r_{TCR}$&$\hat r_{MKTCR}$\\\hline
       	    0.7&3&3.002(2$|$1)&2.993(7$|$0)&2.997(3$|$0)&3.010(1$|$4)&3.042(0$|$11)\\
      	    0.65&3&3.002(3$|$3)&2.990(10$|$0)&2.996(5$|$1)&3.032(1$|$10)&3.064(1$|$17)\\
       	    0.6&3&3.007(7$|$5)&2.982(17$|$0)&2.988(13$|$1)&3.045(4$|$15)&3.077(4$|$22)\\
       	    0.55&3&3.015(14$|$7)&2.963(37$|$0)&2.989(22$|$3)&3.040(10$|$12)&3.097(9$|$28)\\
       	    0.5&3&3.014(37$|$18)&2.909(92$|$2)&2.958(64$|$9)&3.087(23$|$32)&3.180(20$|$54)\\
       	    0.45&3&2.973(74$|$17)&2.824(169$|$3)&2.896(114$|$6)&3.093(58$|$43)&3.248(43$|$79)\\
       	    0.4&3&2.934(147$|$23)&2.714(284$|$1)&2.805(213$|$6)&3.121(102$|$55)&3.340(83$|$105)\\	
  	\bottomrule[1.2pt]		
  	\end{tabular*}\\
  \end{center}
\end{table*}
\begin{table*}[hbpt]
  	\begin{center}
  		\addtolength{\tabcolsep}{11pt}
  		\small
  		\caption{Simulation results for Scenario   $\Bb\mathbf{4}$: $r=3,\theta=1,\rho=0.5,\beta=0.2,J={\rm max}\{10, N/20\},N=T=100$, $(\bF_t^\top,\bv_t^\top)\sim \mathcal{N}(\bf{0},\bI_{N+r})$. Effects of the choice of $k_{{\rm max}}$.}\label{tab:8}
    \renewcommand{\arraystretch}{1.55}
  		\begin{tabular*}{16cm}{lllllll}
  			\toprule[1.2pt]
       		$k_{{\rm max}}$&$r$&$\hat r_{GR}$&$\hat r_{ER}$&$\hat r_{MKER}$&$\hat r_{TCR}$&$\hat r_{MKTCR}$\\\hline
       	    8&3&3.000(0$|$0)&3.000(0$|$0)&3.000(0$|$0)&3.003(0$|$1)&3.003(0$|$3)\\
      	    12&3&3.000(0$|$0)&3.000(0$|$0)&3.000(0$|$0)&3.000(0$|$0)&3.012(0$|$3)\\
       	    16&3&3.000(0$|$0)&3.000(0$|$0)&3.000(0$|$0)&3.006(0$|$1)&3.011(0$|$2)\\
       	    20&3&2.999(1$|$0)&2.998(2$|$0)&2.999(1$|$0)&3.014(0$|$3)&3.030(0$|$6)\\
       	    25&3&3.001(0$|$1)&3.001(0$|$1)&3.001(0$|$1)&3.011(0$|$3)&3.016(0$|$4)\\
       	    30&3&3.005(0$|$2)&3.001(0$|$1)&3.001(0$|$1)&3.017(0$|$5)&3.043(0$|$10)\\
  	\bottomrule[1.2pt]		
  	\end{tabular*}\\
  \end{center}
\end{table*}
\begin{table*}[hbpt]
  	\begin{center}
  		\addtolength{\tabcolsep}{10pt}
  		\small 	
  		\caption{Simulation results for  Scenario  $\Bb\mathbf{5}$: $r=2,\theta=1,\rho=0.5,\beta=0.2,k_{{\rm max}}=8,J={\rm max}\{10, N/20\},N=T=100$, $(\bF_t^\top,\bv_t^\top)\sim\mathcal{N}(\bf{0},\Db)$, $\Db$ is $(N+r)\times(N+r)$ diagonal with $\mathrm{D}_{ii}=1,i\ne 1;\mathrm{D}_{11}=SNR$, $SNR$ from 1 to 20. Effects of dominant factor with two factors.}\label{tab:9}
    \renewcommand{\arraystretch}{1.55}
  		\begin{tabular*}{16.5cm}{lllllll}
  			\toprule[1.2pt]
       		$SNR$&$r$&$\hat r_{GR}$&$\hat r_{ER}$&$\hat r_{MKER}$&$\hat r_{TCR}$&$\hat r_{MKTCR}$\\\hline
       	    1&2&2.000(0$|$0)&2.000(0$|$0)&2.000(0$|$0)&2.000(0$|$0)&2.008(0$|$3)\\
      	    3&2&1.999(1$|$0)&1.967(33$|$0)&1.995(5$|$0)&2.000(0$|$0)&2.005(0$|$1)\\
       	    7&2&1.965(35$|$0)&1.463(537$|$0)&1.855(145$|$0)&1.999(1$|$0)&1.997(3$|$0)\\
       	    10&2&1.907(93$|$0)&1.179(821$|$0)&1.692(308$|$0)&1.994(6$|$0)&1.989(11$|$0)\\
       	    15&2&1.824(176$|$0)&1.039(961$|$0)&1.445(555$|$0)&1.996(5$|$1)&1.995(6$|$1)\\
       	    20&2&1.774(226$|$0)&1.005(995$|$0)&1.267(733$|$0)&1.994(6$|$0)&1.987(14$|$1)\\
  	\bottomrule[1.2pt]		
  	\end{tabular*}\\
  \end{center}
\end{table*}			
\begin{table*}[hbpt]
  	\begin{center}
  		\addtolength{\tabcolsep}{2.8pt}
  		\small \caption{Simulation results for Scenario $\Cb\mathbf{2}$: $r=3,\theta=6,\rho=0.5,\beta=0.2,k_{{\rm max}}=8,J={\rm max}\{10, N/20\}$, $(\bF_t^\top,\bv_t^\top)\sim t_3(\bf{0},\bI_{N+r})$. 	Effects of all weak factors.}\label{tab:10}
    \renewcommand{\arraystretch}{1.55}
  		\vspace{0.2cm}
  		\begin{tabular*}{16.5cm}{llllllll}
  			\toprule[1.2pt]
      		$N$&$T$&$r$&$\hat r_{GR}$&$\hat r_{ER}$&$\hat r_{MKER}$&$\hat r_{TCR}$&$\hat r_{MKTCR}$\\\hline
       	   100&100&3&3.939(446$|$493)&3.192(563$|$364)&6.109(89$|$872)&4.098(424$|$520)&6.527(53$|$922)\\
      	    125&125&3&3.345(508$|$393)&2.696(619$|$281)&5.051(211$|$686)&3.441(491$|$411)&5.523(153$|$762)\\
       	    150&150&3&2.918(537$|$319)&2.457(632$|$229)&3.982(288$|$482)&3.024(520$|$341)&4.470(222$|$567)\\
       	    175&175&3&2.660(547$|$272)&2.282(636$|$191)&3.241(306$|$284)&2.779(522$|$297)&3.655(234$|$379)\\
       	    200&200&3&2.668(506$|$256)&2.309(585$|$173)&2.971(230$|$150)&2.751(485$|$274)&3.202(167$|$200)\\
       	    225&225&3&2.627(491$|$245)&2.321(577$|$180)&2.979(165$|$117)&2.720(464$|$267)&3.160(120$|$158)\\
       	    250&250&3&2.612(500$|$243)&2.322(568$|$174)&2.977(177$|$115)&2.716(480$|$265)&3.180(132$|$167)\\
       	    275&275&3&2.685(463$|$256)&2.374(541$|$178)&2.965(122$|$77)&2.756(442$|$272)&3.090(96$|$114)\\
       	    300&300&3&2.584(475$|$217)&2.285(549$|$154)&3.027(105$|$96)&2.660(455$|$233)&3.174(79$|$136)\\
  	\bottomrule[1.2pt]		
  	\end{tabular*}\\
  \end{center}
\end{table*}
\begin{table*}[hbpt]
  	\begin{center}
  		\addtolength{\tabcolsep}{6pt}
  		\small \caption{Simulation results for  Scenario $\Cb\mathbf{3}$: $r=3,\theta=1,\rho=0.5,\beta=0.2,k_{{\rm max}}=8,J={\rm max}\{10, N/20\},N=T=150$, $(\bF_t^\top,\bv_t^\top)\sim t_3(\bf{0},\Db)$, $\Db$ is $(N+r)\times(N+r)$ diagonal with $\mathrm{D}_{ii}=1,i\ne 3;\mathrm{D}_{33}=SNR$, $SNR$ from 0.7 to 0.4. Effects of strong and weak factors.}
  		\vspace{0.2cm}\label{tab:11}
    \renewcommand{\arraystretch}{1.55}
  		\begin{tabular*}{16cm}{lllllll}
  			\toprule[1.2pt]
       		$SNR$&$r$&$\hat r_{GR}$&$\hat r_{ER}$&$\hat r_{MKER}$&$\hat r_{TCR}$&$\hat r_{MKTCR}$\\\hline
       	    0.7&3&3.077(22$|$102)&2.937(70$|$56)&3.000(0$|$0)&3.142(5$|$123)&3.002(0$|$2)\\
      	    0.65&3&3.049(41$|$104)&2.925(86$|$65)&2.999(1$|$0)&3.129(19$|$135)&3.004(0$|$3)\\
       	    0.6&3&3.045(54$|$113)&2.855(137$|$56)&2.997(3$|$0)&3.140(28$|$148)&3.001(1$|$1)\\
       	    0.55&3&3.003(77$|$107)&2.832(149$|$62)&2.991(10$|$1)&3.104(44$|$142)&3.004(1$|$4)\\
       	    0.5&3&2.994(95$|$113)&2.806(181$|$63)&2.976(24$|$0)&3.125(52$|$150)&2.992(8$|$0)\\
       	    0.45&3&2.982(128$|$124)&2.713(270$|$58)&2.940(62$|$2)&3.081(85$|$156)&2.987(17$|$4)\\
       	    0.4&3&2.870(216$|$103)&2.597(364$|$45)&2.838(164$|$2)&2.996(168$|$135)&2.947(70$|$12)\\	
  	\bottomrule[1.2pt]		
  	\end{tabular*}\\
  \end{center}
\end{table*}
		
	\end{appendices}	

\newpage



\end{document}